\documentclass[a4paper,11pt]{article}
\pdfoutput=1 

\usepackage{jheppub} 

\usepackage[T1]{fontenc} 

\usepackage{amsmath,amsfonts,amsbsy,amssymb,array,accents,dsfont}
\usepackage{enumerate,latexsym,graphicx}
\usepackage{xcolor,tikz}
\usepackage{tcolorbox}

\usepackage{subfigure}
\usepackage{stmaryrd}

\newcommand{\beq}{\begin{equation}}
\newcommand{\eeq}{\end{equation}}
\newcommand{\bea}{\begin{eqnarray}}
\newcommand{\eea}{\end{eqnarray}}

\newcommand{\vep}{\varepsilon}

\newcommand{\der}{\partial}

\newcommand{\nn}{\nonumber}
\newcommand{\<}{\langle}
\renewcommand{\>}{\rangle}

\def \sl{\text{sl}}

\usepackage{braket}
\usepackage{tikz}
\usetikzlibrary{matrix,calc,positioning,decorations.markings,decorations.pathmorphing,decorations.pathreplacing}
\usetikzlibrary{arrows,cd}
\usepackage{bbding}
\usetikzlibrary{positioning}
\tikzset{>=stealth}

\newcommand{\del}{\partial}

\newcommand{\qqquad}{\;, \quad\qquad}  

\newcommand{\SSS}{\mathbb{S}}

\newcommand{\RR}{\mathbb{R}}

\newcommand{\TT}{\mathbb{T}}

\newcommand{\R}{\ensuremath{\mathbb{R}}}

\newcommand{\w}{\omega}

\usepackage{color}

\usepackage[normalem]{ulem}  

\definecolor{darkred}{rgb}{0.6,0,0}
\definecolor{darkblue}{rgb}{0,0,0.6}

\newcommand{\be}{\begin{equation}}
\newcommand{\ee}{\end{equation}}

\usepackage[bbgreekl]{mathbbol}
\usepackage{amsfonts}
\DeclareSymbolFontAlphabet{\mathbb}{AMSb} 
\DeclareSymbolFontAlphabet{\mathbbl}{bbold} 

\title{\boldmath A proof for string three-point functions in AdS$_3$
}

\author[a]{Davide Bufalini,}
\author[b,c
]{Sergio Iguri,}
\author[d]{and Nicolas Kovensky}

\affiliation[a]{Mathematical Sciences and STAG Research Centre,
University of Southampton, Southampton SO17 1BJ, United Kingdom.}
\affiliation[b]{Instituto de Astronomía y Física del Espacio (IAFE) - CONICET and Facultad de Ciencias Exactas y Naturales, Universidad de Buenos Aires, Ciudad Universitaria, 1428 Buenos Aires, Argentina.}
\affiliation[c]{Mathematics with Computer Science Program, Guangdong Technion - Israel Institute of Technology, 515063 Shantou, Guangdong, People's Republic of China.}
\affiliation[d]{Institut de Physique Th\'eorique, Universit\'e Paris Saclay, CEA, CNRS, Orme des Merisiers, 91191 Gif-sur-Yvette CEDEX, France.}

\emailAdd{d.bufalini@soton.ac.uk}
\emailAdd{siguri@iafe.uba.ar}
\emailAdd{nicolas.kovensky@ipht.fr}

\abstract{Correlation functions of the SL(2,$\R$)-WZW model involving spectrally flowed vertex operators are notoriously difficult to compute. An explicit integral expression for the corresponding three-point functions was recently conjectured in \cite{Dei:2021xgh}. In this paper, we provide a proof for this conjecture. For this, we extend the methods of \cite{Iguri:2022eat} based on the so-called SL(2,$\R$) series identifications, which relate vertex operators belonging to different spectral flow sectors. We also highlight the role of holomorphic covering maps in this context. Our results constitute an important milestone for proving this instance of the AdS$_3$/CFT$_2$ holographic duality at finite 't Hooft coupling. 
}

\begin{document} 
\maketitle
\flushbottom

\section{Introduction}

In most instances of the AdS/CFT duality, the bulk theory is approximated by supergravity, due to the notorious difficulty in performing stringy computations. The case of string theory in AdS$_3$ with pure NS fluxes is a notable exception. The dynamics of closed strings propagating on this background can be described at the worldsheet level in terms of the Wess-Zumino-Witten (WZW) model built upon the universal cover of SL$(2,\RR)$. This model is believed to be exactly solvable, hence providing a concrete scenario in which the AdS/CFT duality might be proven at finite 't Hooft coupling.  

The worldsheet CFT, being both Lorentzian and non-compact, enjoys a number of unusual features. Perhaps the most important one is the non-trivial action of the so-called spectral flow outer-automorphisms of the affine sl(2,$\R$)$_k$ algebra. Spectral flow plays a central role in the determination of the string spectrum and partition function \cite{Maldacena:2000hw,Maldacena:2000kv}. In a semi-classical description of long string states, the spectral flow charge $\w$ can be thought of as a winding number around the asymptotic boundary of AdS$_3$. However, it is not a conserved quantity since the associated circle becomes contractible in the AdS$_3$ interior. 

Spectral flow also introduces important complications, especially concerning the computation of worldsheet correlation functions \cite{Maldacena:2001km}. Indeed, while being Virasoro primaries, spectrally flowed vertex operators are \textit{not} affine primaries. Their operator product expansions (OPEs) with the conserved currents become increasingly complicated with growing $\w$, and contain many unknown terms. Hence, some conventional techniques of two-dimensional CFT can not be applied directly. 

The interest in correlation functions of the SL(2,$\R$)-WZW model is explained by their important holographic applications, especially in the context of superstring theory in AdS$_3 \times \SSS^3 \times \TT^4$ (or K3) \cite{Giveon:1998ns,Kutasov:1999xu,Giribet:2007wp,Gaberdiel:2007vu}, and for the dynamics of light probes in black hole microstates \cite{Martinec:2017ztd,Martinec:2018nco,Martinec:2019wzw,Martinec:2020gkv,Martinec:2022okx,Bufalini:2021ndn,Bufalini:2022wyp,Bufalini:2022wzu}. Worldsheet vertex operators are functions of the worldsheet coordinate $z$, and of an additional continuous parameter $x$, which plays the role of the holomorphic coordinate on the conformal boundary. Hence, $z$-integrated correlators are identified with $n$-point functions of local operators in the holographic CFT. Although the precise definition of the latter remains elusive, a concrete albeit perturbative proposal was put forward recently in 
\cite{Eberhardt:2021vsx,Balthazar:2021xeh}.
%
The generators of the spacetime Virasoro algebra are in one-to-one correspondence with those of the $\sl(2,\RR)_k$ algebra, and the worldsheet quantum numbers determine the conformal weight $h$ of dual CFT operators \cite{Giveon:1998ns}. 
Suppressing the anti-holomorphic quantities, we denote the spectrally-flowed worldsheet vertex operators in the so-called $x$-basis as $V_{jh}^{\w}(x,z)$, where $j$ is the SL$(2,\R)$ spin, while  $h$ is the spacetime weight. 

In this paper, we consider genus-zero worldsheet correlators of spectrally flowed vertex operators. 
Recently, progress in the computation of three- and four-point functions has been made in \cite{Eberhardt:2019ywk,Dei:2021xgh,Dei:2021yom}, thanks to the systematic use of the constraints deriving from global and local symmetries. In particular, the latter imply a set of complicated linear recursion relations among correlators involving different assignments of spectral flow charges. These are difficult  to derive in general. However, in \cite{Dei:2021xgh} it has been shown that they can be recast in the form of partial differential equations thanks to the introduction of the so-called `$y$-variable'. In this new basis, a vertex operator is now defined as a coherent superposition of states with different spacetime weights, denoted as $V_{j}^{\w}(x,y,z)$. The authors of \cite{Dei:2021xgh} were able to infer a closed-form expression for all three-point functions 
\begin{equation}
\label{eq: three point}
    \braket{
    V_{j_1}^{\w_1}(x_1;y_1;z_1)
    V_{j_2}^{\w_2}(x_2;y_2;z_2)
    V_{j_3}^{\w_3}(x_3;y_3;z_3)} \, . 
\end{equation}
This proposal can be understood intuitively from the existence, in some cases, of a holomorphic mapping of the worldsheet to the AdS$_3$ boundary, which has branching points of order $\w_i$ at each of the corresponding insertions. The same authors provided a closed-form expression for four-point functions in a follow-up work \cite{Dei:2021yom}. These proposals satisfy a number of highly non-trivial consistency checks. However, in both cases, these expressions were derived on a case-by-case basis, and for a finite set of sufficiently low spectral flow charges.

In this work, we provide a proof for the conjecture of \cite{Dei:2021xgh} concerning the three-point functions \eqref{eq: three point}. Our methods rely heavily on the so-called `series identifications' of SL(2,$\R$), originally formulated in \cite{Fateev,Maldacena:2000hw}, which constitute a set of isomorphisms among affine modules of the $\sl(2,\R)_k$ algebra with adjacent spectral flow charges. The corresponding identities at the level of $y$-basis operators were obtained recently in \cite{Iguri:2022eat}. This allows us to derive the differential equations satisfied by (and the resulting $y$-dependence of) all spectrally flowed three-point functions, including those that do not admit a holomorphic cover.

The structure of the paper is as follows. 
In section \ref{sec: review and conjecture}, we establish our conventions 
and review the derivation of the recursion relations satisfied by the SL(2,$\R$) correlation functions of \cite{Eberhardt:2019ywk}. We also  introduce the $y$-basis operators of \cite{Dei:2021xgh} together with their conjecture for three-point functions. 
Our main results are presented in section \ref{sec: proof}, where derive the partial differential equations satisfied by all $y$-basis three-point functions, and show that all the corresponding solutions are compatible with the proposal of \cite{Dei:2021xgh}. 
In particular, generic odd parity correlators are considered in section \ref{sec:oddcases}, while even parity correlators are obtained in section \ref{sec:evenparitycases}.  
Edge cases and correlators with unflowed insertions are treated in sections \ref{sec: edge cases} and \ref{sec:unflowed}, respectively. Finally, in section \ref{sec: Normalization} we fix the $y$-independent normalisation factors, following the arguments presented in \cite{Iguri:2022eat}. 
We conclude with some discussions and outlook for future work in section \ref{sec: discussion}.

\section{Conventions and brief review of the conjecture} 
\label{sec: review and conjecture}

Let us consider the bosonic SL(2,$\R$)-WZW model at level $k>3$. In this section, we introduce the spectrum and the associated vertex operators. We also briefly review the analysis of \cite{Eberhardt:2019ywk} leading to the recursion relations satisfied by the correlation functions of the model, and present the conjecture put forward in \cite{Dei:2021xgh} for spectrally flowed three-point functions.

\subsection{Vertex operators} 

 We focus mainly on the holomorphic sector. The conserved currents satisfy 
\begin{equation}
    J^a(z)J^b(w) \sim \frac{\eta^{ab} k/2}{(z-w)^2} + \frac{ f^{ab}_{\phantom{ab}c} J^c(w)}{z-w}  \, ,
    \label{OPEjSL2}
\end{equation}
with  $a=+,-,3$, $
\eta^{+-} =-2\eta^{33} =  2$, $f^{+-}_{\phantom{+-}3}=-2$ and $    f^{3+}_{\phantom{3+}+}=-
f^{3-}_{\phantom{3-}-}=1$. In the so-called $x$-basis, vertex operators are denoted as $V^{\w}_{jh}(x,z)$. They depend both on the worldsheet coordinate $z$ and on $x$, the coordinate on the boundary of AdS$_3$ associated with the holographic CFT. Moreover, $h$ and $j$ denote the spacetime weight and the SL(2,$\R$) spin, respectively. The worldsheet conformal weight is
\begin{equation}
    \Delta = -\frac{j(j-1)}{k-2} - h\w + \frac{k}{4}\w^2\,,
\end{equation}
while $\w \geq 0 $ is the spectral flow charge. 

Two types of states must be taken into account \cite{Maldacena:2000hw}. 
For operators defined upon unflowed states in the continuous representations ${\cal{C}}_j^\alpha$, the relevant quantum numbers are  
\begin{equation}
    j \, \in \, \frac{1}{2} + i \R \quad \text{and} 
    \quad m = \alpha \pm n \, ,\quad   \text{with} 
    \quad \alpha \in [0,1)\quad \text{and}   \quad n \in \mathbb{N}_0\, ,  
\end{equation}
where $m$ stands for the spin projection of the corresponding unflowed state.
On the other hand, for the unflowed discrete highest/lowest-weight representations ${\cal{D}}_j^\pm$ we have 
\begin{equation}
    \frac{1}{2} < j < \frac{k-1}{2} \, \quad \text{and} \quad  
     m = \pm (j + n) \, ,\quad \text{with} \quad j \in \R \quad \text{and}   \quad
    n \in \mathbb{N}_0. 
\end{equation}
In both cases we have $h = m + \frac{k}{2}\w$ when $\w>0$, while $h=j$ for $\w=0$. 

The vertex operators are defined by means of their OPEs with the currents: 
\begin{subequations}
\begin{eqnarray}
   J^+(w)  V^{\w}_{jh}(x,z) & = & 
   \sum_{n=1}^{\w+1} \frac{\left(J_{n-1}^+ 
    V_{jh}^\w\right) (x,z) }{(w-z)^n}
    +\cdots \, , \\ [1ex]
    J^3(w)  V^{\w}_{jh}(x,z) & = & 
    x \sum_{n=2}^{\w+1} \frac{\left(J_{n-1}^+ 
    V_{jh}^\w\right) (x,z) }{(w-z)^n}
    +\frac{\left(J_{0}^3 
    V_{jh}^\w\right) (x,z)  
      }{(w-z)}  + \cdots
    \, , \\ [1ex]
    J^-(w)  V^{\w}_{jh}(x,z) & = &  
    x^2 \sum_{n=2}^{\w+1} \frac{\left(J_{n-1}^+ 
    V_{jh}^\w\right) (x,z) }{(w-z)^n}
    +\frac{\left(J_{0}^- 
    V_{jh}^\w\right) (x,z)  
      }{(w-z)}  + \cdots
      \, , 
\end{eqnarray}
\label{JVxOPE}
\end{subequations}
where the ellipsis indicates higher order terms in $(w-z)$. Unflowed vertex operators will be denoted by $V_j(x,z)$. The zero modes act as differential operators in $x$, 
\begin{subequations}
\begin{eqnarray}
    \left(J_{0}^+V_{jh}^\w\right) (x,z) &=& \der_x V_{jh}^\w (x,z) \, , 
    \label{J0+dx}\\[1ex]
    \left(J_{0}^3V_{jh}^\w\right) (x,z) &=& (x\der_x+h) V_{jh}^\w (x,z) \, , \\[1ex]
    \left(J_{0}^-V_{jh}^\w\right) (x,z) &=& (x^2\der_x + 2hx) V_{jh}^\w (x,z) \, ,
\end{eqnarray}    
\label{diffopsx}
\end{subequations}
while 
\begin{equation}
    \left(J_{\pm \w}^\pm V_{jh}^\w\right) (x,z) = \left[h-\frac{k}{2} \w \pm (1-j)\right] V_{j,h\pm 1}^\w (x,z)  \, . 
    \label{JwVx}
\end{equation}
Importantly, in terms of the currents
\begin{equation}
J^+(x,z)=J^+(z) 
\,, \quad 
    J^3(x,z) = J^3(z) - x J^+(z) \, , \quad J^-(x,z) = J^-(z) - 2 x J^3(z)+x^2 J^+(z)
\label{defJx}
\end{equation} 
we get 
\begin{subequations}
\begin{eqnarray}
    J^3(x,w)  V^{\w}_{jh}(x,z) & = & 
    \frac{h 
      }{(w-z)} V_{jh}^\w(x,z) + \cdots
    \, , \label{J3xOPE}\\ [1ex]
    J^-(x,w)  V^{\w}_{jh}(x,z) & = &  
    (w-z)^{\w-1} \left(J_{-w}^-V_{jh}^\w\right) (x,z) + \cdots
    \, . 
    \label{JmxOPE}
\end{eqnarray}
\end{subequations}

An alternative (equivalent) definition was given recently in \cite{Iguri:2022eat}, based on \cite{Maldacena:2001km}. This is to be understood as a point-splitting procedure between the corresponding unflowed vertex and the so-called generalized spectral flow operator $V_{\frac{k}{2},\frac{k}{2}\w}^{\w-1} (x,z)$, and reads 
\begin{equation}
    V_{jh}^{\w}(x,z) = 
    \lim_{\vep,\bar{\vep}\to 0}  \vep^{m \w}
    \bar{\vep}^{\bar{m} \w}
    \int d^2y \,  y^{j-m-1}  \bar{y}^{j-\bar{m}-1}  V_{j}(x+y,z+\vep) V_{\frac{k}{2},\frac{k}{2}\w}^{\w-1} (x,z) \, .
    \label{proposalWxbasis2}
\end{equation}
We will come back to this shortly in section \ref{sec: conjecture}.  

\subsection{Recursion relations among correlators}
\label{sec:recursions}

We now discuss the correlation functions of the model. It was shown in \cite{Eberhardt:2019ywk} that they must satisfy a set of recursion relations. Let us briefly review how this works for three-point functions of the form 
\begin{equation}
\label{defF}
    F =\left\langle \prod_{j=1}^3 V_{j_j h_j}^{\w_j}(x_j,z_j)\right\rangle. 
\end{equation}
We define 
\begin{equation}
    F_n^i =\left\langle 
    \left(J_n^+V_{j_i h_i}^{\w_i}\right)(x_i,z_i)
    \prod_{j\neq i} V_{j_j h_j}^{\w_j}(x_j,z_j)\right\rangle,  
    \label{defFin}
\end{equation}
so that, in particular, 
\begin{equation}
    F_0^i = \der_{x_i} F\, ,
\end{equation}
and 
\begin{equation}
    F_{\w_i}^i = \left(h_i-\frac{k}{2} \w_i + 1-j_i\right) \left\langle V_{j_i, h_i+1}^{\w_i}(x_i,z_i) \prod_{j\neq i} V_{j_j h_j}^{\w_j}(x_j,z_j)\right\rangle\,,
    \label{Fwi}
\end{equation}
as implied by Eqs.~\eqref{J0+dx} and \eqref{JwVx} respectively. Note that all $F^i_n$ with $n=1,\dots, \w_i-1$ are, in principle, unknown. 

By using the OPEs in Eq.~\eqref{JVxOPE}, one finds that correlators involving a current insertion can be expanded as
\begin{subequations}
\begin{eqnarray}
    \left\langle J^+(z) \prod_{j=1}^3 V_{j_j h_j}^{\w_j}(x_j,z_j)\right\rangle &=& \sum_{i=1}^3 \left[
    \frac{\der_{x_i} F}{z-z_i} + \sum_{n=1}^{\w_i} 
    \frac{F_{n}^i}{(z-z_i)^{n+1}}
    \right] + \cdots \, , \\[1ex]
    \left\langle J^3(z) \prod_{j=1}^3 V_{j_j h_j}^{\w_j}(x_j,z_j)\right\rangle &=& \sum_{i=1}^3 \left[
    \frac{(h_i+x_i\der_{x_i}) F}{z-z_i} + \sum_{n=1}^{\w_i} 
    \frac{x_i F_{n}^i}{(z-z_i)^{n+1}}
    \right] + \cdots\, , \\[1ex]
    \left\langle J^-(z) \prod_{j=1}^3 V_{j_j h_j}^{\w_j}(x_j,z_j)\right\rangle &=& \sum_{i=1}^3 \left[
    \frac{(2h_i x_i + x_i^2\der_{x_i} )F}{z-z_i} + \sum_{n=1}^{\w_i} 
    \frac{x_i^2 F_{n}^i}{(z-z_i)^{n+1}}
    \right] + \cdots. 
\end{eqnarray}
\end{subequations}
Combining these expressions, we get 
\begin{equation}
\label{Gjdef}
    G_j(z) \equiv \left\langle J^-(x_j,z) \prod_{l=1}^3 V_{j_l h_l}^{\w_l}(x_l,z_l)\right\rangle = 
    \sum_{i\neq j} \left[
    \frac{(2h_i x_{ij} + x_{ij}^2\der_{x_i} )F}{z-z_i} + \sum_{n=1}^{\w_i} 
    \frac{x_{ij}^2 F_{n}^i}{(z-z_i)^{n+1}}
    \right] + \cdots
\end{equation}
where $x_{ij} = x_i - x_j$. 

On the other hand, Eq.~\eqref{JmxOPE} imposes stringent restrictions on the behavior of $G_j(z)$ when $z$ is close to $z_j$. More precisely, in this regime, we must have 
\begin{equation}
     (z-z_j)^{1-\w_j}G_j(z) =  
    \left(h_j-\frac{k}{2} \w_j - 1+j_j\right) \left\langle V_{j_j, h_j-1}^{\w_j}(x_j,z_j) \prod_{i\neq j} V_{j_i h_i}^{\w_i}(x_i,z_i)\right\rangle + \cdots, 
    \label{Gjconditions}
\end{equation}
where we have used \eqref{JwVx}. As it turns out, the regularity of $(z-z_j)^{1-\w_j}G_j(z)$ at $z=z_j$  as implied by \eqref{Gjconditions} for all $j=1,2,3$ provides enough information to solve for all the unknown $F_n^i$ in terms of $F$, its $x_i$-derivatives, and $F_{\w_i}^i$. 
Upon inserting the resulting expressions into Eq.~\eqref{Gjconditions}, one obtains a set of complicated linear relations between correlators involving spectrally flowed vertex operators with consecutive spacetime weights and their $x_i$ derivatives. Note that the latter are under control since the global Ward identities uniquely fix the $x_i$-dependence of all such correlators.   

\subsection{The conjectured solution}
\label{sec: conjecture}

The non-trivial OPEs in Eqs.~\eqref{JVxOPE} render the computation of correlation functions involving spectrally flowed insertions quite complicated. It will be useful to work with somewhat unusual linear combinations of the operators $V_{jh}^{\w}(x,z)$ with the same $j$ and $\w$ but different values of $h$. These are the $y$-basis operators $V_j^\w (x,y,z)$ introduced in \cite{Dei:2021xgh}. In \cite{Iguri:2022eat}, these were shown  to be precisely the integrands on the RHS of Eq.~\eqref{proposalWxbasis2}, namely 
\begin{equation}
    V_j^\w (x,y,z) \equiv \lim_{\vep,\bar{\vep}\to 0}  |\vep|^{2j \w} V_{j}(x+y\vep^\w,z+\vep) V_{\frac{k}{2}, \frac{k}{2}\w}^{\w-1} (x,z)\,.
    \label{defVWxyz}
\end{equation}
This can be understood directly from Eq.~\eqref{proposalWxbasis2} above. Indeed, upon changing variables $y \to y \vep^\w$ this can be re-written as 
\begin{equation}
    V_{jh}^{\w}(x,z) = 
    \int d^2y \,  y^{j-m-1}  \bar{y}^{j-\bar{m}-1} 
    V_j^\w (x,y,z)\, ,
    \label{proposalWxybasis}
\end{equation}
which coincides with the so-called $y$-transform of \cite{Dei:2021xgh}. More details can be found in \cite{Dei:2021xgh,Iguri:2022eat}.

The OPEs of $y$-basis operators with the conserved currents are 
\begin{subequations}
\begin{eqnarray}
   J^+(w)  V^{\w}_{j}(x,y,z) & = & 
   \sum_{n=1}^{\w+1} \frac{\left(J_{n-1}^+ 
    V_{jh}^\w\right) (x,y,z) }{(w-z)^n}
    +\cdots \, , \\
    J^3(x,w)  V^{\w}_{j}(x,y,z) & = & 
    \frac{y \der_y + j + \frac{k}{2}\w
      }{(w-z)} \, V_{j}^\w(x,y,z) + \cdots
    \, , \\[1ex]
    J^-(x,w)  V^{\w}_{j}(x,y,z) & = &  
    (w-z)^{\w-1} \left(J_{-w}^-V_{j}^\w\right) (x,y,z) + \cdots
    \, . 
\end{eqnarray}
\label{JVxyOPE}
\end{subequations}
While the zero modes still act as in \eqref{diffopsx}, the main motivation for using the $y$-variable is that $J_{\pm \w}^\pm$ act as differential operators in $y$. More precisely, we have  
\begin{subequations}
\label{JwVy}
\begin{eqnarray}
    \left(J_{\w}^+V_{j}^\w\right) (x,y,z) &=& \der_y V_{j}^\w (x,y,z) \, , 
    \\[1ex]
    \left(J_{-\w}^- V_{j}^\w\right) (x,y,z) &=& (y^2\der_y + 2jy) V_{j}^\w (x,y,z) \, .
    \label{diffopsyMinus}
\end{eqnarray}    
\label{diffopsy}
\end{subequations}
This allows one to re-write the recursion relations derived formally in the previous section as differential equations for correlators of the form  
\begin{equation}
    F_y 
    \equiv \langle V_{j_1 }^{\w_1}(x_1,y_1,z_1)V_{j_2 }^{\w_2}(x_2,y_2,z_2)V_{j_3 }^{\w_3}(x_3,y_3,z_3)\rangle\, .
    \label{V1V2V3xy}
\end{equation}
The $x$-basis correlators in Eq.~\eqref{defF} can be obtained from these by means of \eqref{proposalWxybasis}\footnote{The procedure is slightly different for flowed discrete and continuous states. The former arise as residues from poles of the integrand in \eqref{proposalWxybasis} around $y=0$ or $y= \infty$, depending on whether the corresponding unflowed vertex operator belongs to a lowest- or highest-weight representation. For the flowed continuous states one must integrate over the full complex plane.}. This integration procedure can be complicated, and was only carried out explicitly in \cite{Dei:2021xgh} for a subset of cases.

In order to discuss the structure of these differential equations and their solutions, it will be useful to make use of the conformal invariance on the worldsheet and boundary CFT to fix $x_1 = z_1=0$, $x_2 = z_2=1$ and $x_3 = z_3=\infty$, and consider 
\begin{equation}
    \hat{F}_y = 
    \langle V_{j_1 }^{\w_1}(y_1)V_{j_2 }^{\w_2}(y_2)V_{j_3 }^{\w_3}(y_3)\rangle\equiv \langle V_{j_1 }^{\w_1}(0,y_1,0)V_{j_2 }^{\w_2}(1,y_2,1)V_{j_3 }^{\w_3}(\infty,y_3,\infty)\rangle\,
    .\label{Vw1w2w301infty}
\end{equation}  
The latter is related to the original correlator in Eq.~\eqref{V1V2V3xy} by 
\begin{align}
\begin{aligned}
   &\langle V_{j_1 }^{\w_1}(x_1,y_1,z_1)V_{j_2 }^{\w_2}(x_2,y_2,z_2)V_{j_3 }^{\w_3}(x_3,y_3,z_3)\rangle =  \frac{x_{21}^{h^0_3-h^0_1-h^0_2}
    x_{31}^{h^0_2-h^0_1-h^0_3}
    x_{32}^{h^0_1-h^0_2-h^0_3}}{
    z_{21}^{\Delta^0_1+\Delta^0_2-\Delta^0_3}
    z_{31}^{\Delta^0_1+\Delta^0_3-\Delta^0_2}
    z_{32}^{\Delta^0_2+\Delta^0_3-\Delta^0_1}}
    \times \,  \\[1ex]
    &      \left\< V_{j_1}^{\w_1} \left(0,y_1 \frac{x_{32} z_{21}^{\w_1}z_{31}^{\w_1}}{x_{21}x_{31}z_{32}^{\w_1}},0\right)V_{j_2}^{\w_2} \left(1,
    y_2 \frac{x_{31} z_{21}^{\w_2}z_{32}^{\w_2}}{x_{21}x_{32}z_{31}^{\w_2}}
    ,1\right)V_{j_3}^{\w_3}\left(\infty,
    y_3 \frac{x_{21} z_{31}^{\w_3}z_{32}^{\w_3}}{x_{31}x_{32}z_{21}^{\w_3}},\infty\right)\right\> ,
\end{aligned}
\label{ybasisx1x2x3fixing}
\end{align}
where 
\begin{equation}
    h^0_i = j_i + \frac{k}{2}\w_i
    \, , \qquad \Delta^0_i 
    = - \frac{j_i(j_i-1)}{k-2} - j_i \w_i - \frac{k}{4}\w_i^2\, .
\end{equation}

Except for a certain subfamily of correlators which will be discussed below,  spectrally flowed $y$-basis three-point functions and their associated differential equations were studied in \cite{Dei:2021xgh} on a case-by-case basis. This was done for sufficiently low values of the spectral flow charges $\w_i$, thus leading the authors to conjecture a general solution for the $y$-dependence of these correlators. The proposed expressions (not including the right-movers, and up to an overall normalization constant to be discussed below) read as follows. 
For \textit{odd parity} correlators, namely when $\w_1+\w_2+\w_3 \in 2 \mathbb{Z}+1$, one has 
\begin{align}
\left\langle V^{\w_1}_{j_1}(0, y_1, 0) \,  V^{\w_2}_{j_2}(1, y_2, 1) \, V^{\w_3}_{j_3}(\infty, y_3, \infty) \right\rangle = 
X_{123}^{\frac{k}{2}-j_1-j_2-j_3} \prod_{i=1}^3 X_i^{-\frac{k}{2}+j_1+j_2+j_3-2j_i} \ ,
\label{3pt-odd-parity}
\end{align}
while for the \textit{even parity} case, i.e.~when $\w_1+\w_2+\w_3 \in 2 \mathbb{Z}$, 
\begin{align}
\left\langle V^{\w_1}_{j_1}(0, y_1, 0) \,  V^{\w_2}_{j_2}(1, y_2, 1) \, V^{\w_3}_{j_3}(\infty, y_3, \infty) \right\rangle = X_\emptyset^{j_1+j_2+j_3-k}\prod_{i<\ell} X_{i \ell}^{j_1+j_2+j_3-2j_i-2j_\ell} \  .
\label{3pt-even-parity}
\end{align}
Here, for any subset $I \subset \{ 1,2,3 \}$, 
\be 
X_I(y_1,y_2,y_3)\equiv \sum_{i \in I:\ \varepsilon_i=\pm 1} P_{\boldsymbol{\w}+\sum_{i \in I} \varepsilon_i e_{i}} \prod_{i\in I} y_i^{\frac{1-\varepsilon_i}{2}} \ . 
\label{X_I-3pt}
\ee
The spectral flow parameters are chosen as $\boldsymbol{\w}=(\w_1, \w_2, \w_3)$, while $
e_1 = (1,0,0)$, $e_2 = (0,1,0)$ and $e_3 = (0,0,1)$.
The numbers $P_{\boldsymbol{\w}}$ are defined as  
\be 
P_{\boldsymbol{\w}} = 0 \qquad \text{for} \qquad \sum_j \w_j < 2 \max_{i=1,2,3} \w_i \quad \text{or}\quad \sum_i \w_i \in 2\mathds{Z}+1
\ee
and 
\be
P_{\boldsymbol{\w}} =S_{\boldsymbol{\w}} \frac{G\left(\frac{-\w_1+\w_2+\w_3}{2} +1\right) G\left(\frac{\w_1-\w_2+\w_3}{2} +1\right) G\left(\frac{\w_1+\w_2-\w_3}{2} +1\right) G\left(\frac{\w_1+\w_2+\w_3}{2}+1\right)}{G(\w_1+1) G(\w_2+1) G(\w_3+1)}  \ , 
\label{Pw-definition}
\ee
otherwise, where $G(n)$ is the Barnes G function
\be 
G(n)=\prod_{i=1}^{n-1} \Gamma(i)
\label{barnesG}
\ee
for positive integer values, while $S_{\boldsymbol{\w}}$ is a phase depending on $\boldsymbol{\w} \bmod 2$. For more details, see \cite{Dei:2021xgh}. 

Regarding the overall constants, their precise form was also conjectured in \cite{Dei:2021xgh} and later proven in \cite{Iguri:2022eat}. These structure constants are  
\begin{equation}
\label{consdei}
    C_{\boldsymbol{\w}}(j_1,j_2,j_3) = \left\{\begin{array}{cc}
    C(j_1,j_2,j_3),     & \text{if} \quad  \w_1 + \w_2 + \w_3 \in 2\mathbb{Z}, \\
    {\cal{N}}(j_1) C\left(\frac{k}{2}-j_1,j_2,j_3 \right),     & \qquad  \text{if} \quad  \w_1 + \w_2 + \w_3 \in 2\mathbb{Z}+1, 
    \end{array} \right.
\end{equation}  
where $C(j_1,j_2,j_3)$ are the structure constants of the unflowed three-point functions defined in terms of Barnes double Gamma functions in \cite{Teschner:1999ug,Maldacena:2001km}. Finally, ${\cal{N}}(j_1)$ is defined in terms of the reflection coefficient appearing the unflowed two-point functions, namely 
\begin{equation}
    {\cal{N}}(j) = \sqrt{\frac{B(j)}{B\left(
    \frac{k}{2}-j\right)}} \, ,     \label{Ndef}
\end{equation}
with\footnote{To be precise, $\nu$ is actually a free parameter of the model, which essentially plays the role of the string coupling. Here we simply reproduce the value originally advocated in \cite{Teschner:1999ug}. For a related discussion, see \cite{Dabholkar:2007ey,Eberhardt:2021vsx}. We thank A.~Dei and L.~Eberhardt for pointing this out.} 
\begin{equation}
    B(j)=\frac{2j-1}{\pi}
    \frac{\Gamma[1-b^2(2j-1)]}{\Gamma[1+b^2(2j-1)]} \, \nu^{1-2j} \, , \quad 
    \nu = \frac{\Gamma[1-b^2]}{\Gamma[1+b^2]}
    \, , \quad  b^2 = (k-2)^{-1}
     \label{defBj} \, .
\end{equation}

As shown in \cite{Dei:2021xgh}, the proposal given in Eqs.~\eqref{3pt-odd-parity}-\eqref{consdei} passes a number of non-trivial consistency checks, including  bosonic exchange symmetry and  reflection symmetry for continuous states. 
However, no general expression for the $y$-basis differential equations is known. Hence, so far, this conjecture remains to be proven. 

In the remainder of the paper, we prove that this solution is indeed correct. In doing so, we highlight the role of holomorphic covering maps. Although these maps appear to be essential for the study of four-point functions \cite{Dei:2021yom}, we show that they also play a key role in the present context. Furthermore, when treating cases where there is no well-defined covering map available, we will make use of the so-called series identifications for spectrally flowed vertex operators constructed upon states belonging to the discrete representations of SL(2,$\R$) \cite{Fateev,Iguri:2022eat}.

\section{The proof for $y$-basis three-point functions}
\label{sec: proof}

In this section we prove the conjecture put forward in \cite{Dei:2021xgh}. 
It was shown in \cite{Maldacena:2001km} that all non-vanishing spectrally flowed three-point functions in the SL(2,$\R$) model must satisfy the following condition: 
\begin{equation}
    \w_i+\w_j \geq \w_k-1 \qquad \forall \,\, 
    i \neq j \neq k. 
    \label{fusionrules}
\end{equation}
We first consider the subfamily of odd parity correlators for which an associated holomorphic covering map exists \cite{Eberhardt:2019ywk,Dei:2021xgh}. We then show how to compute all remaining non-vanishing correlators with three non-trivial spectral flow charges. This includes even parity correlators, and also those we denote as edge correlators. The latter correspond to correlators for which either the inequality in \eqref{fusionrules} for odd parity assignments or the analogous inequality for even parity cases saturate. These need to be treated with special care. Finally, we also discuss correlators with unflowed insertions and the overall normalizations.    

\subsection{Holomorphic covering maps and odd parity correlators}
\label{sec:oddcases}

For concreteness, and with no loss of generality, we take $\w_3$ to be the largest spectral flow charge, i.e.~$\w_3\geq \w_{1,2}$. Let us consider correlators satisfying
\begin{equation}
    \w_1+\w_2+\w_3 \in 2\mathbb{Z}+1 
    \qqquad \w_1+\w_2 > \w_3 - 1 
    \qqquad \w_i\geq 1\,, \,\, \forall \, i. \label{conditionsmapwi}
\end{equation}
It was shown in \cite{Lunin:2000yv} that there exists a  unique holomorphic covering map $\Gamma[\w_1,\w_2,\w_3](z) \equiv \Gamma(z)$ from the worldsheet to the AdS$_3$ boundary such that  
\begin{equation}
\Gamma(z) \sim x_i + a_i (z-z_i)^{\w_i} + \cdots\qquad \text{when} \quad 
z\sim z_i\,,
\label{coveringmapexp}
\end{equation}
where the ellipsis indicates higher order terms in $(z-z_i)$.  
This is a rational function which approaches a constant $\Gamma_\infty$ as $z\to \infty$. One can show that it develops 
$N$ single poles, counted by the Riemann-Hurwitz formula \begin{equation}
    N = \frac{1}{2}\left(\w_1+\w_2+\w_3-1\right).
\end{equation}  
The  coefficients $a_i$ appearing in Eq.~\eqref{coveringmapexp} take the form 
\begin{equation}
    a_i =  
    \left(
    \begin{array}{c}
        \frac{\w_1+\w_2+\w_3-1}{2}  \\
        \frac{-\w_1+\w_2+\w_3-1}{2}
    \end{array}\right)
   \left(
    \begin{array}{c}
        \frac{-\w_1+\w_2-\w_3-1}{2} \\
        \frac{\w_1+\w_2-\w_3-1}{2}
    \end{array}\right)^{-1}
    \frac{x_{i,i+1}x_{i+2,i}}{
    x_{i+1,i+2}}
    \left(\frac{z_{i+1,i+2}}{
    z_{i,i+1}z_{i+2,i}}\right)^{\w_i}\,, 
    \label{coveringmapcoeffs}
\end{equation}
where the subscripts are understood mod 3.
Note that the last two factors in \eqref{coveringmapcoeffs} simplify to 1 upon setting $x_1=z_1= 0$, $x_2=z_2= 1$ and $x_3=z_3= \infty$. We will use the same notation $a_i$ for the resulting purely numerical coefficients. 

We now derive the differential equations satisfied by $y$-basis three-point functions satisfying \eqref{conditionsmapwi}. Although the presentation here is slightly different, this was already done in \cite{Eberhardt:2019ywk}.  Consider the operator $J^-(\Gamma(z),z)$, where we use the notation of Eq.~\eqref{JmxOPE}, namely 
\begin{equation}
    J^-(\Gamma(z),z) = J^-(z)-2\Gamma(z) J^3(z)
    +\Gamma^2(z) J^+(z) \, .
    \label{JGamma}
\end{equation}
In order to obtain the recursion relations for the correlators under consideration, we compute following  the contour integral: 
\begin{equation}
    \oint_{z_i} \frac{dz}{(z-z_i)^{\w_i}} \left\langle 
    J^-(\Gamma(z),z)\,
    V_{j_1}^{\w_1}(x_1,y_1,z_1)
    V_{j_2}^{\w_2}(x_2,y_2,z_2)
    V_{j_3}^{\w_3}(x_3,y_3,z_3) \right\rangle \, .
    \label{ointJgammazi}
\end{equation}
Similarly to what was done in Sec.~\ref{sec:recursions} above, we do this in two different ways. First, we note that, near $z_i$, we have 
\begin{equation}
    J^-(\Gamma(z),z) = J^-(x_i,z) 
    - 2 a_i (z-z_i)^{\w_i} J^3(x_i,z)
    +a_i^2 (z-z_i)^{2\w_i} J^+(z) + \cdots.
\end{equation}
Hence, by using \eqref{JVxyOPE} we find that 
\begin{equation}
   \eqref{ointJgammazi} =  \left[ 
    \left(2j_i y_i + y_i^2 \der_{y_i}\right) 
    -2 a_i  \left( j_i+\frac{k}{2}\w_i + y_i \der_{y_i}\right)
    +a_i^2 \der_{y_i}
    \right] 
    F_y\, ,
\label{ointJgammazi1}
\end{equation}
where $F_y$ was defined in Eq.~\eqref{V1V2V3xy}. 
On the other hand, using \eqref{JGamma} together with the OPEs in Eq.~\eqref{JVxOPE}, one finds that 
\begin{align}
    \begin{aligned}
&   \left\langle 
    J^-(\Gamma(z),z)\,
     V_{j_1}^{\w_1}(x_1,y_1,z_1)
    V_{j_2}^{\w_2}(x_2,y_2,z_2)
    V_{j_3}^{\w_3}(x_3,y_3,z_3)  \right\rangle =  \\ 
    & \qquad \qquad = \sum_{j=1}^{3} 
    \left\{
    -\frac{2 [\Gamma(z)-x_j](y_j \der_{y_j} + j_j + \frac{k}{2}\w_j)}{z-z_j} F_y + \sum_{n=1}^{\w_j}
    \frac{[\Gamma(z)-x_j]^2}{(z-z_j)^{n+1}} F_{y,n}^i
    \right\} \,,
    \label{ointJgammazi2} 
    \end{aligned}
\end{align}
where $F_{y,n}^i$ stands for the $y$-basis analogues of the $F_{n}^i$ defined in Eq.~\eqref{defFin}. As discussed in \cite{Eberhardt:2019ywk}, the RHS of \eqref{ointJgammazi2} is a rational function of $z$ which, as implied by the constraint equations, has zeros of order $\w_i-1$ at all $z_i$. It also has double poles at the $N$ simple poles of $\Gamma(z)$, and further goes to zero as $z^{-2}$ for $z\to \infty$ due to the global Ward identities. This implies that it must be proportional to the derivative of the covering map, namely    
\begin{equation}
    \left\langle 
    J^-(\Gamma(z),z)\,
     V_{j_1}^{\w_1}(x_1,y_1,z_1)
    V_{j_2}^{\w_2}(x_2,y_2,z_2)
    V_{j_3}^{\w_3}(x_3,y_3,z_3)  \right\rangle = \alpha \, \der \Gamma(z) \, , 
\end{equation}
where $\alpha$ must be independent of $z$. This coefficient was also computed in  \cite{Eberhardt:2019ywk}. When working in the  $y$-basis it takes the following form: 
\begin{equation}
    \alpha = -\frac{1}{N}\sum_{j=1}^3\left(
    (y_j-a_j)\der_{y_j} + j_j + \frac{k}{2}\w_j \right) F_y
    \, . 
\end{equation}
This allows us to provide an alternative expression for the contour integral \eqref{ointJgammazi}. Indeed, the behavior of the covering map near the insertion points showcased in \eqref{coveringmapexp} implies that 
\begin{equation}
\eqref{ointJgammazi}=      -\frac{a_i \w_i}{N}\sum_{j=1}^3\left(
    (y_j-a_j)\der_{y_j} + j_j + \frac{k}{2}\w_j \right) F_y
    \, .
    \label{RHSodd}
\end{equation}

By combining the results in Eqs.~\eqref{ointJgammazi1} and \eqref{RHSodd}, and further fixing the insertion points as in \eqref{Vw1w2w301infty},  one obtains the following differential equations:   
\begin{equation}
 \left\{    (y_i-a_i)^2 \der_{y_i} 
    +2 j_i (y_i-a_i) +
    \frac{a_i \w_i}{N}\left[ 
    \sum_{j=1}^3\left(
    (y_j-a_j)\der_{y_j} + j_j \right) -\frac{k}{2} \right] \right\} \hat{F}_y = 0,
\label{recursionywithmap}
\end{equation}
for $i=1,2,3$. This was derived originally in this form in \cite{Dei:2021xgh}\footnote{Note that we have corrected a couple of typos in their presentation.}. 
In this way, the use of the covering map and its derivative allows one to avoid dealing with the cumbersome unknowns discussed in Sec.~\ref{sec:recursions}. 

The system of equations encoded in \eqref{recursionywithmap} uniquely fixes the dependence of the corresponding correlators on $y_1$, $y_2$ and $y_3$. Up to some overall normalization, which will be discussed in section \ref{sec: Normalization} below, the solution of \eqref{recursionywithmap} is
\begin{eqnarray}
    \langle V_{j_1}^{\w_1}(y_1)
    V_{j_2}^{\w_2}(y_2)
    V_{j_3}^{\w_3}(y_3) \rangle &=& (y_1-a_1)^{j_2+j_3-j_1-\frac{k}{2}}
    (y_2-a_2)^{j_3+j_1-j_2-\frac{k}{2}}
    (y_3-a_3)^{j_1+j_2-j_3-\frac{k}{2}}  \nn \\
    && \quad \times \left[
    \sum_{\vep_{1,2,3}=\pm 1} 2 \frac{N_1^{\vep_1,\vep_2,\vep_3}}{N}\prod_{i=1}^3
    a_i^{\frac{\vep_i+1}{2}} y_i^{\frac{\vep_i-1}{2}}
    \right]^{\frac{k}{2}-j_1-j_2-j_3}. 
    \label{oddfinal1}
\end{eqnarray}
An equivalent, perhaps simpler expression for the same solution is given by
\begin{align}
\langle V_{j_1}^{\w_1}(y_1)
    V_{j_2}^{\w_2}(y_2)
    V_{j_3}^{\w_3}(y_3) \rangle & = (y_1-a_1)^{-2j_1}
    (y_2-a_2)^{-2j_2}
    (y_3-a_3)^{-2j_3} \label{oddfinal2} \\
    &\quad \times \left(
    \w_1\frac{y_1+a_1}{y_1-a_1}
    +\w_2\frac{y_2+a_2}{y_2-a_2}
    +\w_3\frac{y_3+a_3}{y_3-a_3} -1
    \right)^{\frac{k}{2}-j_1-j_2-j_3}. \nn 
\end{align}  
Note that this formulation makes the bosonic exchange symmetry manifest.

At first sight, this expression might seem very different from that in Eq.~\eqref{3pt-odd-parity}. The connection comes from the fact that while the $P_{\boldsymbol{\w}}$ defined in \eqref{Pw-definition} are somewhat complicated, their ratios are actually much simpler. For instance, consider the $X_i$ term appearing in \eqref{3pt-odd-parity}. From Eqs.~\eqref{X_I-3pt} and \eqref{Pw-definition}, up to an overall sign, we find that 
\begin{equation}
    X_i = P_{\boldsymbol{\w}-e_i} y_i + P_{\boldsymbol{\w}+e_i} = P_{\boldsymbol{\w}-e_i} \left(y_i+\frac{P_{\boldsymbol{\w}+e_i}}{P_{\boldsymbol{\w}-e_i}}\right) = 
    P_{\boldsymbol{\w}-e_i} \left(y_i-a_i\right) \, .
\end{equation}
A similar result holds for $X_{123}$, showing that the $y$-dependence of the expression in \eqref{oddfinal2} is consistent with that of Eq.~\eqref{3pt-odd-parity}. 

As shown in Eq.~\eqref{oddfinal2}, $y$-basis three-point functions diverge whenever a variable $y_i$ approaches the corresponding coefficient $a_i$. Thus, the $a_i$ are very special points in the $y$-plane, which signal the existence of an appropriate holomorphic covering map. An even more extreme situation takes place in the tensionless limit, which corresponds to $k=3$ in the bosonic language \cite{Giribet:2018ada,Gaberdiel:2018rqv,Eberhardt:2018ouy,Eberhardt:2019ywk}. There, spectrally flowed correlators are non-vanishing \textit{only} when $y_i = a_i$ for all $i$. The corresponding recursion relations \eqref{recursionywithmap} are then satisfied provided\footnote{In the supersymmetric case, the RNS formalism breaks down for the tensionless theory. It was shown in  \cite{Eberhardt:2018ouy}, using the so-called hybrid formalism, that in this case only vertex operators with $j=\frac{1}{2}$ are allowed.}  $j_1+j_2+j_3=\frac{3}{2}$. It would be interesting to fully understand the relation between $y$-variables and covering map coefficients in the general case.

We end this section by noting that the same discussion can not be applied directly to odd parity correlators saturating the inequality in Eq.~\eqref{fusionrules}, i.e.~those with $\w_3 = \w_1+\w_2+1$. These will be discussed in detail in section \ref{sec: edge cases} below. Correlators with unflowed insertions are further considered in section \ref{sec:unflowed}. 

\subsection{Series identification and even parity correlators}
\label{sec:evenparitycases}
We now prove that the conjecture of \cite{Dei:2021xgh} also holds for correlators satisfying 
\begin{equation}
    \w_1 + \w_2 + \w_3  \, \in 2  \mathbb{Z}\qqquad
    \w_1+\w_2 > \w_3 
    \qqquad \w_i\geq 1\,, \,\, \forall \, i.
    \label{wiEvenCases}
\end{equation}
These include all non-vanishing spectrally flowed three-point functions for which the total spectral flow charge is even, except for the edge cases where $\w_3 = \w_1 + \w_2$ and/or some of the vertex operators are unflowed, which will be treated separately. 

When the $\w_i$ are as in \eqref{wiEvenCases}, it is not possible to construct a covering map such as the one used in the previous section. Thus, one might wonder if differential equations similar to those in \eqref{recursionywithmap} can be deduced in this context. Indeed, the corresponding recursion relations have only been obtained on a case-by-case basis and for sufficiently low values of the spectral flow charges \cite{Dei:2021xgh}. 

Nevertheless, we observe that the procedure outlined in Sec.~\ref{sec:recursions} guarantees that, provided the system  is compatible, and once the $F_n^i$ have been solved for, the resulting $y$-basis recursions must take the form 
\begin{equation}
    \left[y_i(y_i\der_{y_i} + 2j_i) + \sum_{j=1}^3(A_{ij} y_j -B_{ij})\der_{y_j} 
    + C_i \right]\langle V_{j_1}^{\w_1}(y_1)
    V_{j_2}^{\w_2}(y_2)
    V_{j_3}^{\w_3}(y_3) \rangle = 0 \, , 
    \label{GenEqsEven}
\end{equation}
where $A_{ij}$, $B_{ij}$ and $C_i$ are some numerical constants to be determined, which depend on the spins $j_i$ and the charges $\w_i$. 
The rationale behind the structure of Eq.~\eqref{GenEqsEven}, which is the $y$-basis version of Eq.~\eqref{Gjconditions}, goes as follows. First, note that, upon using Eq.~\eqref{diffopsyMinus}, the operator $y_i(y_i\der_{y_i} + 2j_i)$ is identified with  the RHS of \eqref{Gjconditions}. Second, recall that the recursion relations were derived by expressing the OPEs of the vertex operators with the conserved currents in terms of the unknowns coming from the action of the modes of $J^+(z)$, see Eq.~\eqref{Gjdef}. This implies that the term involving the action of $J^-(x,z)$ does not mix with the rest. By using the Möbius-fixed expression for the three-point function in Eq.~\eqref{ybasisx1x2x3fixing}, one can see that the terms in the recursion relations involving unknowns and $x$-derivatives of the correlator are mapped to operators of the schematic form $y \del_y$ and $\del_y$, as well as $y$-independent multiplicative factors. In Eq.~\eqref{GenEqsEven} we have allowed for generic coefficients $A_{ij}$, $B_{ij}$ and $C_{i}$ in front of the corresponding contributions.

Moreover, we note that the way in which these equations are derived only depends on the values of the spectral flow charges involved in a given correlator. In other words, for a given set of $\w_i$, the recursion relations are independent of whether the vertex operators involved belong to spectrally flowed discrete or continuous representations.   
These two observations will allow us to obtain all $y$-basis differential equations associated with even parity correlators in closed form. 

As it turns out, even and odd parity cases are not completely disconnected. For the discrete representations, affine modules in spectral flow sectors with one unit of difference in the spectral flow charge are identifiable. For $y$-basis operators the corresponding series identifications read \cite{Iguri:2022eat}
\begin{equation}
 V_j^\w(x,y=0,z) =  {\cal{N}}(j)
 \lim_{y\to \infty}  y^{k-2j} V_{\frac{k}{2}-j}^{\w+1}(x,y,z),
 \label{seriesidybasis1}
\end{equation}
and 
\begin{equation}
 \lim_{y\to \infty}  y^{2j} V_j^\w(x,y,z) = {\cal{N}}(j)
 V_{\frac{k}{2}-j}^{\w-1}(x,y=0,z), 
 \label{seriesidybasis2}
\end{equation} 
where ${\cal{N}}(j)$ was defined in Eq.~\eqref{Ndef}.  
It was shown recently in \cite{Iguri:2022eat} that, assuming the $y$-dependence proposed in \cite{Dei:2021xgh} for all three-point functions, these relations fix the $y$-basis structure constants in terms of the unflowed ones, which can be found in  \cite{Maldacena:2001km}. Here we show that Eqs.~\eqref{seriesidybasis1} and \eqref{seriesidybasis2} are actually much more powerful: they allow us to fix the $y$-dependence as well. More explicitly, we use them to derive all unknown coefficients $A_{ij}$, $B_{ij}$, and $C_i$ appearing in \eqref{GenEqsEven}. 

By means of Eqs.~\eqref{seriesidybasis1} and \eqref{seriesidybasis2}, we find that all even parity correlators can be related to (at least) three \textit{different} situations where a covering map satisfying \eqref{coveringmapexp} and \eqref{coveringmapcoeffs} exists. Explicitly, given ($\w_1,\w_2,\w_3$) satisfying \eqref{wiEvenCases}, it follows that the adjacent assignments $(\w_1+1,\w_2,\w_3)$, $(\w_1,\w_2+1,\w_3)$ and $(\w_1,\w_2,\w_3-1)$ satisfy all conditions in \eqref{conditionsmapwi}. Let us denote the corresponding covering maps as follows:
\begin{equation}
    \Gamma_1^+ \equiv \Gamma[\w_1+1,\w_2,\w_3] \qqquad 
    \Gamma_2^+ \equiv \Gamma[\w_1,\w_2+1,\w_3] \qqquad 
    \Gamma_3^- \equiv \Gamma[\w_1,\w_2,\w_3-1]\,.
    \label{3maps1}
\end{equation}
Then, the relations \eqref{seriesidybasis1} and \eqref{seriesidybasis2} provide the following set of identities: 
\begin{subequations}
\begin{eqnarray}
&&\langle V_{j_1}^{\w_1}(0)
    V_{j_2}^{\w_2}(y_2)
    V_{j_3}^{\w_3}(y_3) \rangle = \lim_{y_1\to \infty} y_1^{k-2j_1} {\cal{N}}(j_1) 
    \langle V_{\frac{k}{2}-j_1}^{\w_1+1}(y_1)
    V_{j_2}^{\w_2}(y_2)
    V_{j_3}^{\w_3}(y_3) \rangle, \\[1ex]
&&\langle V_{j_1}^{\w_1}(y_1)
    V_{j_2}^{\w_2}(0)
    V_{j_3}^{\w_3}(y_3) \rangle = \lim_{y_2\to \infty} y_2^{k-2j_2} {\cal{N}}(j_2) 
    \langle V_{j_1}^{\w_1}(y_1)
    V_{\frac{k}{2}-j_2}^{\w_2+1}(y_2)
    V_{j_3}^{\w_3}(y_3) \rangle, \\[1ex]
&&\lim_{y_3\to \infty} y_3^{2j_3}\langle        V_{j_1}^{\w_1}(y_1)
    V_{j_2}^{\w_2}(y_2)
    V_{j_3}^{\w_3}(y_3) \rangle = {\cal{N}}(j_3) 
    \langle V_{j_1}^{\w_1}(y_1)
    V_{j_2}^{\w_2}(y_2)
    V_{\frac{k}{2}-j_3}^{\w_3-1}(0) \rangle.
\end{eqnarray}
\label{3mapsCorrs1}
\end{subequations}
Having excluded the even edge cases, the same holds for the adjacent assignments $(\w_1-1,\w_2,\w_3)$, $(\w_1,\w_2-1,\w_3)$ and $(\w_1,\w_2,\w_3+1)$, the corresponding maps being 
\begin{equation}
    \Gamma_1^- \equiv \Gamma[\w_1-1,\w_2,\w_3] \qqquad 
    \Gamma_2^- \equiv \Gamma[\w_1,\w_2-1,\w_3] \qqquad 
    \Gamma_3^+ \equiv \Gamma[\w_1,\w_2,\w_3+1]\,. 
       \label{3maps2}
\end{equation}
These lead to 
\begin{subequations}
\begin{eqnarray}
&&\lim_{y_1\to \infty} y_1^{2j_1}\langle V_{j_1}^{\w_1}(y_1)
    V_{j_2}^{\w_2}(y_2)
    V_{j_3}^{\w_3}(y_3) \rangle = {\cal{N}}(j_1) 
    \langle V_{\frac{k}{2}-j_1}^{\w_1-1}(0)
    V_{j_2}^{\w_2}(y_2)
    V_{j_3}^{\w_3}(y_3) \rangle \\[1ex]
&&\lim_{y_2\to \infty} y_2^{2j_2}\langle V_{j_1}^{\w_1}(y_1)
    V_{j_2}^{\w_2}(y_2)
    V_{j_3}^{\w_3}(y_3) \rangle = {\cal{N}}(j_2) 
    \langle V_{j_1}^{\w_1}(y_1)
    V_{\frac{k}{2}-j_2}^{\w_2-1}(0)
    V_{j_3}^{\w_3}(y_3) \rangle,  \\[1ex]
&&\langle V_{j_1}^{\w_1}(y_1)
    V_{j_2}^{\w_2}(y_2)
    V_{j_3}^{\w_3}(0) \rangle = \lim_{y_3\to \infty} y_3^{k-2j_3} {\cal{N}}(j_3) 
    \langle V_{j_1}^{\w_1}(y_1)
    V_{j_2}^{\w_2}(y_2)
    V_{\frac{k}{2}-j_3}^{\w_3+1}(y_3) \rangle. 
\end{eqnarray}
\label{3mapsCorrs2}
\end{subequations}

All expressions on the RHS of Eqs.~\eqref{3mapsCorrs1} and \eqref{3mapsCorrs2} are limits of correlators discussed in the previous section. Hence, they must satisfy the appropriate limits of the differential equations given in \eqref{recursionywithmap}. For instance, $\langle V_{\frac{k}{2}-j_1}^{\w_1-1}(0)
V_{j_2}^{\w_2}(y_2) V_{j_3}^{\w_3}(y_3) \rangle$ is annihilated by the differential operators
\begin{align}
    \begin{aligned}
    & y_2 (y_2 \der_{y_2}+2j_2) + (\w_1 - \w_2 - \w_3)^{-1} \big\{ (\w_1 + \w_2 - \w_3)a_2[\Gamma_1^-]^2 \der_{y_2} \\ 
    &  - 2 a_2[\Gamma_1^-] \left[ (\w_1-\w_3)(j_2+y_2 \der_{y_2}) + \w_2(j_3+y_3 \der_{y_3} - a_3[\Gamma_1^-] \der_{y_3} + j_1+j_3) \right]\big\} 
    \end{aligned}
    \label{ex1a}
\end{align}
and
\begin{align}
    \begin{aligned}
    &y_3 (y_3 \der_{y_3}+2j_3) + (\w_1 - \w_2 - \w_3)^{-1} \big\{ (\w_1 - \w_2 + \w_3)a_3[\Gamma_1^-]^2 \der_{y_3} \\ 
    &- 2 a_3[\Gamma_1^-] \left[ (\w_1-\w_2)(j_3+y_3 \der_{y_3}) + \w_3(j_2+y_2 \der_{y_2} - a_2[\Gamma_1^-] \der_{y_2} + j_1+j_2) \right] \big\},  
    \end{aligned}
    \label{ex1b}
\end{align}
where $a_i[\Gamma_1^-]$ denotes the coefficient $a_i$ associated with the map $\Gamma_1^-$. One can obtain analogous equations from the other odd parity correlators involved in \eqref{3mapsCorrs1}. This leads to twelve differential operators, which must coincide with the appropriate limits of those provided in Eqs.~\eqref{GenEqsEven}.  
As an example, in the situation considered above we should match \eqref{ex1a} and \eqref{ex1b} with 
\begin{align}
    \begin{aligned}
    y_2 (y_2 \der_{y_2} + 2j_2) + A_{22} y_2 \der_{y_2}
    + A_{23} y_3 \der_{y_3} - B_{22} \der_{y_2} - B_{23} \der_{y_3} - 2 A_{21} j_1 + C_2, \\
    y_3 (y_3 \der_{y_3} + 2j_3) + A_{33} y_3 \der_{y_3}
    + A_{32} y_2 \der_{y_2} - B_{33} \der_{y_3} - B_{32} \der_{y_2} - 2 A_{31} j_1 + C_3.
    \end{aligned}
\end{align}

After identifying all linearly independent terms in these 12 equations we find a total of 60 conditions.
In the end, 21 of these 60 conditions can be used to solve explicitly for all the coefficients $A_{ij}$, $B_{ij}$ and $C_i$ in \eqref{GenEqsEven}. Consistency of the system demands that the remaining 39 conditions must hold. 
We emphasize that the fact that these are identically satisfied is a highly non-trivial check of the logic behind our proof, and also a remarkable consequence of the identities relating the $a_i$ coefficients of the different covering maps involved.

There are many equivalent ways to write the resulting coefficients. We find that the simplest one is as follows:   
\begin{equation}
\label{AijMatrix}
A=    \left(\begin{array}{ccc}
    \frac{2 (\w_3-\w_2)}{\w_1+\w_2-\w_3} a_1[\Gamma_3^-]     & \frac{2 \w_1}{\w_1+\w_2-\w_3} a_1[\Gamma_3^-]   & 
    \frac{2 \w_1}{\w_1-\w_2+\w_3} a_1[\Gamma_2^-]    \\
    \frac{2 \w_2}{\w_1+\w_2-\w_3} a_2[\Gamma_3^-]        &
    \frac{2 (\w_1-\w_3)}{-\w_1+\w_2+\w_3} a_2[\Gamma_1^-]   &
    \frac{2 \w_2}{-\w_1+\w_2+\w_3} a_2[\Gamma_1^-]   \\
    \frac{2 \w_3}{\w_1-\w_2+\w_3} a_3[\Gamma_2^-]        &
    \frac{2 \w_3}{-\w_1+\w_2+\w_3} a_3[\Gamma_1^-]   &
    \frac{2 (\w_1-\w_2)}{-\w_1+\w_2+\w_3} a_3[\Gamma_1^-]   
    \end{array}\right),
\end{equation}
\begin{equation}
\label{BijMatrix}
B=    \left(\begin{array}{ccc}
    \frac{(\w_1 - \w_2 + \w_2)}{\w_1+\w_2-\w_3} 
    a_1[\Gamma_3^-]^2     & 
    \frac{2 \w_1}{\w_1+\w_2-\w_3} a_1[\Gamma_3^-]a_2[\Gamma_3^-]   & 
    \frac{2 \w_1}{\w_1-\w_2+\w_3} a_1[\Gamma_2^-]a_3[\Gamma_2^-]    \\
    \frac{2 \w_2}{\w_1+\w_2-\w_3} a_1[\Gamma_3^-]a_2[\Gamma_3^-]        &
    \frac{(\w_1 + \w_2 - w_2)}{-\w_1+\w_2+\w_3} a_2[\Gamma_1^-]^2   &
    \frac{2 \w_2}{-\w_1+\w_2+\w_3} a_2[\Gamma_1^-]a_3[\Gamma_1^-]   \\
    \frac{2 \w_3}{\w_1-\w_2+\w_3} a_1[\Gamma_2^-]a_3[\Gamma_2^-]        &
    \frac{2 \w_3}{-\w_1+\w_2+\w_3} a_2[\Gamma_1^-]a_3[\Gamma_1^-]   &
    \frac{(\w_1 - \w_2 + w_2)}{-\w_1+\w_2+\w_3} a_3[\Gamma_1^-]^2   
    \end{array}\right),
\end{equation}
and 
\begin{equation}
\label{CiMatrix}
C=  
   \left(\begin{array}{c}
\frac{4 \w_1 j_3}{\w_1-\w_2+\w_3}a_1[\Gamma_2^-] + 
\frac{2\w_1 (j_2+j_3)+2j_1 (\w_3-\w_2)}{\w_1+\w_2-\w_3}a_1[\Gamma_3^-]\\
\frac{4 \w_2 j_1}{\w_1+\w_2-\w_3}a_2[\Gamma_3^-] - 
\frac{2\w_2 (j_1+j_3)+2j_2 (\w_1-\w_3)}{\w_1-\w_2-\w_3}a_2[\Gamma_1^-]     \\
\frac{4 \w_3 j_1}{\w_1-\w_2+\w_3}a_3[\Gamma_2^-] - 
\frac{2\w_3 (j_1+j_2)+2j_3 (\w_1-\w_2)}{\w_1+\w_2-\w_3}a_3[\Gamma_1^-]
    \end{array}\right).
\end{equation}
Consequently, we can write the differential equations for the even parity correlators as 
\begin{align}
    \begin{aligned}
   &\Big\{ \left(y_1-a_1[\Gamma_3^-]\right)^2 \der_{y_1} + 
    2 j_1 \left(y_1-a_1[\Gamma_3^-]\right) 
    + \frac{2 a_1[\Gamma_3^-] \w_1}{\w_1 + \w_2 - \w_3} \Big[
    (y_1 - a_1[\Gamma_3^-]) \der_{y_1} + j_1  \\
    &\, + \,  (y_2 - a_2[\Gamma_3^-]) \der_{y_2} + j_2
    -(y_3 - a_3[\Gamma_2^-]) \der_{y_3} - j_3
    \Big] \Big\} \, \langle V_{j_1}^{\w_1}(y_1)
    V_{j_2}^{\w_2}(y_2)
    V_{j_3}^{\w_3}(y_3) \rangle = 0\, ,      
    \end{aligned}
    \label{Eq1even}
\end{align}
\begin{align}
    \begin{aligned}
   &\Big\{ \left(y_2-a_2[\Gamma_3^-]\right)^2 \der_{y_2} + 
    2 j_2 \left(y_2-a_2[\Gamma_3^-]\right) 
    + \frac{2 a_2[\Gamma_3^-] \w_2}{\w_1 + \w_2 - \w_3} \Big[
    (y_1 - a_1[\Gamma_3^-]) \der_{y_1} + j_1  \\
    &\, + \,  (y_2 - a_2[\Gamma_3^-]) \der_{y_2} + j_2
    -(y_3 - a_3[\Gamma_1^-]) \der_{y_3} - j_3
    \Big] \Big\} \, \langle V_{j_1}^{\w_1}(y_1)
    V_{j_2}^{\w_2}(y_2)
    V_{j_3}^{\w_3}(y_3) \rangle = 0\,,     
    \end{aligned}
    \label{Eq2even}
\end{align}
and 
\begin{align}
    \begin{aligned}
   &\Big\{ \left(y_3-a_3[\Gamma_1^-]\right)^2 \der_{y_3} + 
    2 j_3 \left(y_3-a_3[\Gamma_1^-]\right) 
    + \frac{2 a_3[\Gamma_1^-] \w_3}{\w_1 - \w_2 - \w_3} \Big[
    (y_1 - a_1[\Gamma_2^-]) \der_{y_1} + j_1  \\
    &\, - \,  (y_2 - a_2[\Gamma_1^-]) \der_{y_2} - j_2
    -(y_3 - a_3[\Gamma_1^-]) \der_{y_3} - j_3
    \Big] \Big\} \, \langle V_{j_1}^{\w_1}(y_1)
    V_{j_2}^{\w_2}(y_2)
    V_{j_3}^{\w_3}(y_3) \rangle = 0 \,.     
    \end{aligned}
    \label{Eq3even}
\end{align}
The structural similarity of these equations with those of the odd case in Eq.~\eqref{recursionywithmap} is striking. This suggest that there must be a way to derive Eqs.~\eqref{Eq1even}, \eqref{Eq2even} and \eqref{Eq3even} directly from the \textit{adjacent} covering maps along the lines of Sec.~\ref{sec:oddcases}. We will not attempt to do this here. 

By solving Eqs.~\eqref{Eq1even}, \eqref{Eq2even} and \eqref{Eq3even}, we find that, up to an overall constant, even parity correlators satisfying \eqref{wiEvenCases} take the form 
\begin{eqnarray}
\label{eq: general even solution}
    \langle V_{j_1}^{\w_1}(y_1)
    V_{j_2}^{\w_2}(y_2)
    V_{j_3}^{\w_3}(y_3) \rangle &= &
    \left(
    1-\frac{y_2}{ a_2[\Gamma_3^+]} -\frac{y_3}{a_3[\Gamma_2^+]}+
    \frac{y_2 y_3}{a_2[\Gamma_3^-] a_3[\Gamma_2^+] }\right)^{j_1-j_2-j_3} \nn \\
    & \times &   \left(
    1-\frac{y_1}{a_1[\Gamma_3^+]}-\frac{y_3}{a_3[\Gamma_1^+]} + \frac{y_1 y_3}{ a_1[\Gamma_3^-]  a_3[\Gamma_1^+]}
    \right)^{j_2-j_3-j_1} \\
    &\times &  \left(
    1-\frac{y_1}{a_1[\Gamma_2^+]} -\frac{y_2}{a_2[\Gamma_1^+]} +\frac{y_1 y_2}{a_1[\Gamma_2^+]a_2[\Gamma_1^-]} \right)^{j_3-j_1-j_2}. \nn
\end{eqnarray}
As in the odd case, one can check that this is consistent with the conjectured expressions in Eq.~\eqref{3pt-even-parity} by using the relations between covering map coefficients and ratios of the numbers $P_{\boldsymbol{\w}}$ defined in \eqref{Pw-definition}. It is also straightforward to see that \eqref{eq: general even solution} matches all relevant limits of the corresponding odd cases in Eq.~\eqref{oddfinal1}, as implied by the different series identifications in Eqs.~\eqref{3mapsCorrs1} and \eqref{3mapsCorrs2}. We also note that, upon using some identities among the $a_i$ coefficients such as $a_1[\Gamma_2^+]a_2[\Gamma_1^-] = a_1[\Gamma_2^-]a_2[\Gamma_1^+]$, the solution in Eq.~\eqref{eq: general even solution} manifestly enjoys bosonic exchange symmetry.

\subsection{Edge cases}
\label{sec: edge cases}

In this subsection, we consider spectral flow assignments lying at the ``edge'' of the inequalities displayed in Eqs.~\eqref{conditionsmapwi} and \eqref{wiEvenCases}. More explicitly, we consider three-point functions with spectral flow charges satisfying either 
\begin{eqnarray}
\label{wiedgecases}
    \w_3 = \w_1+\w_2 \quad \text{or} \quad  \w_3 = \w_1+\w_2 +1\, ,   \qquad \w_i\geq 1 \, , \, \forall \, i\, . 
\end{eqnarray}
Following the nomenclature of the previous sections, we refer to these as the even and odd edge cases, respectively. Note that, according to the fusion rules in Eq.~\eqref{fusionrules}, these include all possibly non-vanishing correlators which were not included in sections \ref{sec:oddcases} and \ref{sec:evenparitycases} above, except for correlators with unflowed insertions, which will be discussed later on. 

The treatment of the edge cases is slightly different from what we have discussed so far. Indeed, the general method based on series identifications employed in Sec.~\ref{sec:evenparitycases} breaks down when considering the even edge cases. We find that many of the coefficients in Eqs.~\eqref{AijMatrix}, \eqref{BijMatrix} and \eqref{CiMatrix} become either divergent or indeterminate when $\w_1 + \w_2=\w_3$, related to the fact that three of the six adjacent maps cease to exist, namely those in \eqref{3maps2}. Although, in principle, the existence of the three maps in \eqref{3maps1} could give enough constraints, in practice one runs into similar problems with divergent or indeterminate coefficients. As for the odd edge cases, it turns out that the associated covering map does not exist. Moreover, the former are only related to even edge cases by the SL(2,$\R$) series identifications. 

We now show that alternative techniques involving current insertions can be used to derive the relevant differential equations satisfied by the $y$-basis edge correlators. 
Some of these equations are easier to derive in the limit where the first two vertex operators collide. As discussed in \cite{Dei:2021xgh}, one can take the vertex operators to be inserted at $(x_1,x_2,x_3) = (0,x,\infty)$, and then consider the limit $x\to 0$. More explicitly, we have  \cite{Dei:2021xgh}
\begin{align}
\label{eq: Mobiusunfix}
    & \braket{
    V^{\w_1}_{j_1}(0,y_1,0)
    V^{\w_2}_{j_2}(x,y_2,1)
    V^{\w_3}_{j_3}(\infty,y_3,\infty)  } \nn\\
     & = x^{-j_1 - j_2 + j_3 + \frac{k}{2}(-\w_1-\w_2 + \w_3)}
     \braket{
     V^{\w_1}_{j_1}\left(0,\frac{y_1}{x},0\right)
     V^{\w_2}_{j_2}\left(1,\frac{y_2}{x},1\right)
     V^{\w_3}_{j_3}\left(\infty,y_3 x,\infty \right)  } \, . 
\end{align}
The three-point functions that remain well-defined in this limit are those for which 
\begin{equation}
\label{wimbasislimit}
    |\w_1 + \w_2 - \w_3| \le 1 \, ,
\end{equation}
including both edge cases in \eqref{wiedgecases}. In these instances, it is possible to obtain the correlator with vertex operators inserted at generic points $(x_1,x_2,x_3)$ from the one in the colliding limit with $(x,x,x_3)$ by means of the global Ward identities. 
We will derive some of the relevant differential equations satisfied by the edge correlators in the above collision limit, and only recover the full correlators at the end. Note that taking $x\to 0$ and $x_3 \to \infty$ precisely corresponds to the limit in which these correlators can be interpreted as $m$-basis correlators of flowed primaries as in \cite{Maldacena:2001km, Cagnacci:2013ufa}, where they were denoted as spectral flow \textit{conserving} and spectral flow \textit{violating} three-point functions, depending on the overall parity of the spectral flow charges. 

To illustrate how this works, let us derive a constraint that will be satisfied by both edge cases. We set $x_1 = x_2 = x$ and consider the integral 
\begin{equation}
\oint_{{\cal{C}}} \braket{
J^3(x,z)
V^{\w_1}_{j_1}(x,y_1,z_1)
V^{\w_2}_{j_2}(x,y_2,z_2)
V^{\w_3}_{j_3}(x_3,y_3,z_3)  }dz \, ,  
\end{equation}
where ${\cal{C}}$ denotes a contour encircling all three insertion points. This trivially vanishes as there is no residue at infinity. On the other hand, the action of the current on the vertex operators at $x$ can be read off directly from Eq.~\eqref{JVxyOPE}, while for the one inserted at $x_3$ we use 
\begin{equation}
    J^3(x,z) =  J^3(x_3,z) + (x_3 - x) J^+(x_3,z) \, ,
\end{equation}
which follows from \eqref{defJx}. We obtain the following differential equation:
\begin{align}
\label{eq:chargeconsfixed}
    \left[
    \sum_{i=1}^3 \left(y_i \der_{y_i} + j_i + \frac{k}{2} \w_i\right) + (x_3-x)\der_{x_3}
    \right] \braket{
V^{\w_1}_{j_1}(x,y_1,z_1)
V^{\w_2}_{j_2}(x,y_2,z_2)
V^{\w_3}_{j_3}(x_3,y_3,z_3)  } =0 \, . 
\end{align}
We now Möbius-fix the worldsheet insertions to $(0,1,\infty)$, and  further consider the limit $(x,x_3)\to (0,\infty)$. The $y_i$ coordinates then get rescaled according to Eq.~\eqref{ybasisx1x2x3fixing}. 
This leads to 
\begin{align}
\begin{aligned}
\label{eq: charge conservation general}
    &\Big[
    y_1 \del_{y_1}   + y_2 \del_{y_2}  - y_3 \del_{y_3} + j_1 + j_2 - j_3  \\ 
    & \qquad + \frac{k}{2} (\w_1 + \w_2 - \w_3)
    \Big] \braket{
V^{\w_1}_{j_1}(0,y_1,0)
V^{\w_2}_{j_2}(0,y_2,1)
V^{\w_3}_{j_3}(\infty,y_3,\infty)  } =0 \, . 
\end{aligned}
\end{align}
We find that Eq.~\eqref{eq: charge conservation general} is the $y$-basis version of the usual charge-conservation equation for $m$-basis three-point functions of spectrally flowed primaries. This holds for all correlators satisfying \eqref{wimbasislimit}, including both edge cases.

\subsubsection{Even edge cases}

In this subsection we focus on the even edge cases, where $\w_3 = \w_1+\w_2$. We will derive the remaining two differential equations by considering correlators with an extra insertion of the $J^-(x,z)$ current multiplied by the appropriate ratio of worldsheet coordinates. This is similar to what was used in \cite{McElgin:2015eho} when discussing the proof of the $m$-basis spectral flow \textit{violation} rules, and also more recently in \cite{Iguri:2022pbp} in the context of the supersymmetric version of this model.

The integral 
\begin{equation}
\label{eq: Jminusinsertedeveneasy}
\oint_{{\cal{C}}} \braket{J^-(x_3,z)  
V^{\w_1}_{j_1}(x_1,y_1,z_1)
V^{\w_2}_{j_2}(x_2,y_2,z_2)
V^{\w_3}_{j_3}(x_3,y_3,z_3)  } \frac{(z-z_1)^{\w_1} (z-z_2)^{\w_2}}{(z-z_3)^{\w_3}} dz 
\end{equation}
vanishes since there is no pole at infinity. Using the OPEs of $J^{-}(x_3,z)$ with the vertex operators, this yields 
\begin{align}
    \Big[ x_{31}^2 \frac{z_{12}^{\w_2}}{z_{13}^{\w_3}} \del_{y_1} & + x_{32}^2 \frac{z_{21}^{\w_1}}{z_{23}^{\w_3}} \del_{y_2} + z_{31}^{\w_1}z_{32}^{\w_2} (y_3^2 \del_{y_3} + 2j_3 y_3)    \Big] \nn \\
    & \braket{ 
    V^{\w_1}_{j_1}(x_1,y_1,z_1)
    V^{\w_2}_{j_2}(x_2,y_2,z_2)
    V^{\w_3}_{j_3}(x_3,y_3,z_3)  }  = 0 \, . 
\end{align}
Proceeding similarly with 
\begin{equation}
    \label{eq: Jminusinsertedevencollision}
    \oint_{{\cal{C}}}  \braket{J^-(x,z) 
    V^{\w_1}_{j_1}(x,y_1,z_1)
    V^{\w_2}_{j_2}(x,y_2,z_2)
    V^{\w_3}_{j_3}(x_3,y_3,z_3)  } \frac{(z-z_3)^{\w_3}}{(z-z_1)^{\w_1}(z-z_2)^{\w_2} }  dz \, , 
\end{equation}
where we have imposed the collision limit mentioned above, we find 
\begin{align}
\Big[\frac{z_{13}^{\w_3}}{z_{12}^{\w_2}} (y_1^2 \del_{y_1} + 2j_1 y_1) & + \frac{z_{23}^{\w_3}}{z_{21}^{\w_2}} (y_2^2 \del_{y_2} + 2j_2 y_2) + \frac{(x-x_3)^2}{z_{31}^{\w_1}z_{32}^{\w_2} } \del_{y_3}   \Big] \nn \\
    & \braket{ 
    V^{\w_1}_{j_1}(x,y_1,z_1)
    V^{\w_2}_{j_2}(x,y_2,z_2)
    V^{\w_3}_{j_3}(x_3,y_3,z_3)  }  = 0 \, . 
\end{align}
We now fix the worldsheet coordinates to $(0,1,\infty)$ while sending $ x \to 0$ and $x_3 \to \infty$, and use Eq.~\eqref{ybasisx1x2x3fixing} for the corresponding rescaling of the $y$ variables. Including the charge conservation condition \eqref{eq: charge conservation general}, the system of differential equations satisfied by the even edge correlator in the collision limit is then
\begin{align}
    \begin{aligned}
    \label{eq: three eq even}
    0 &=    \Big[
    j_1 + j_2 - j_3 
    + y_1 \del_{y_1} + y_2 \del_{y_2}  - y_3 \del_{y_3}
    \Big] \braket{\dots}\, , \\
      0 &=  \Big[  (-1)^{\w_1}\del_{y_1}  + (-1)^{\w_3}\del_{y_2} + (y_3^2 \del_{y_3} + 2j_3 y_3)    \Big] \braket{\dots}  \, , \\
      0 &=  \Big[(-1)^{\w_1} (y_1^2 \del_{y_1} + 2j_1 y_1) + (-1)^{\w_3}(y_2^2 \del_{y_2} + 2j_2 y_2) +  \del_{y_3}   \Big] \braket{\dots} \, .  
    \end{aligned}
\end{align}
where $\braket{\dots}$ stands for $\braket{
V^{\w_1}_{j_1}(0,y_1,0)
V^{\w_2}_{j_2}(0,y_2,1)
V^{\w_3}_{j_3}(\infty,y_3,\infty)  }$.
Note that only the second equation in \eqref{eq: three eq even} remains valid away from the collision limit.

Up to an overall $y$-independent constant, the general solution of the system \eqref{eq: three eq even} can be written as follows: 
\begin{align}
    &\braket{
    V^{\w_1}_{j_1}(0,y_1,0)
    V^{\w_2}_{j_2}(0,y_2,1)
    V^{\w_3}_{j_3}(\infty,y_3,\infty)  }  \\
    & = ((-1)^{\w_1} y_1 -  (-1)^{\w_3}y_2)^{j_3 -j_1 - j_2 }
    (1 + (-1)^{\w_3}y_2 y_3 )^{j_1 - j_2 - j_3} 
    (1 + (-1)^{\w_1}y_1 y_3)^{ j_2  - j_1  - j_3} \, . \nn 
\end{align}
This matches the result of \cite{Dei:2021xgh}, see their Eq.~(5.37b).

As mentioned above, for more  general values of the insertion points the corresponding three-point functions follow from the global Ward identities. As it turns out, we can infer the result in a heuristic way by looking at the general expression given in Eq.~\eqref{eq: general even solution}. Indeed, one can verify that, upon setting $\w_3 = \w_1 + \w_2$, the coefficients $a_1(\Gamma_3^+), a_2(\Gamma_3^+)$ and $a_2(\Gamma_1^-)$ diverge, which is a manifestation of the fact that the associated covering maps do not exist. Since all other coefficients remain finite, we obtain 
\begin{eqnarray}
\label{eq: even edge solution}
& \langle 
    V_{j_1}^{\w_1}(y_1)
    V_{j_2}^{\w_2}(y_2)
    V_{j_3}^{\w_3}(y_3) \rangle  & =   \left(
    1 + (-1)^{\w_1}\frac{(\w_1 + \w_2-1)!}{(\w_1-1)!\w_2!}y_3
    +
    (-1)^{\w_3}y_2 y_3 \right)^{j_1-j_2-j_3} 
    \nn \\
    &\times & \hspace{-2cm} 
     \left(
    1 + (-1)^{\w_1+1} \frac{(\w_1 + \w_2-1)!}{\w_1!(\w_2-1)!}
    y_3 + (-1)^{\w_1}y_1 y_3
    \right)^{j_2-j_3-j_1} \\
    &\times &  \hspace{-2cm}
   \left(
    1 + (-1)^{\w_1+1} \frac{\w_1 !\w_2!}{(\w_1+\w_2)!}((-1)^{\w_1} y_1 - (-1)^{\w_3} y_2) \right)^{j_3-j_1-j_2} .\nn 
\end{eqnarray}
Up to the normalisation, to be discussed below, this precisely matches the conjecture \eqref{3pt-even-parity}.

\subsubsection{Odd edge cases}

We now turn to the odd edge cases, where $\w_3 = \w_1+\w_2+1$. Since the procedure is analogous to what we just described we will skip some of the intermediate steps. 

In addition to \eqref{eq: charge conservation general}, we find two differential equations by considering  contour integrals very similar to that in Eq.~\eqref{eq: Jminusinsertedeveneasy}. We first take  
\begin{equation}
\label{eq: Jminusinsertedodd1}
\oint_{{\cal{C}}} \braket{J^-(x_3,z)  V^{\w_1}_{j_1}(x_1,y_1,z_1)
V^{\w_2}_{j_2}(x_2,y_2,z_2)
V^{\w_3}_{j_3}(x_3,y_3,z_3)  } \frac{(z-z_1)^{\w_1+1} (z-z_2)^{\w_2}}{(z-z_3)^{\w_3}} dz \, , 
\end{equation}
which again vanishes due to the absence of a residue at infinity. The same holds for \begin{equation}
\label{eq: Jminusinsertedodd1}
\oint_{{\cal{C}}} \braket{J^-(x_3,z)  V^{\w_1}_{j_1}(x_1,y_1,z_1)
V^{\w_2}_{j_2}(x_2,y_2,z_2)
V^{\w_3}_{j_3}(x_3,y_3,z_3)  } \frac{(z-z_1)^{\w_1} (z-z_2)^{\w_2+1}}{(z-z_3)^{\w_3}} dz \, . 
\end{equation}
Hence, we find the following system of differential equations: 
\begin{align}
    \begin{aligned}
    \label{eq: three eq odd}
    0 & =    \Big[
    j_1 + j_2 - j_3 - \frac{k}{2}
    + y_1 \del_{y_1} + y_2 \del_{y_2}  - y_3 \del_{y_3}
    \Big] \braket{\dots}  \, , \\
    0 & = \Big[ (-1)^{\w_3} \del_{y_2} +  (y_3^2 \del_{y_3} + 2 j_3 y_3  )   \Big] \braket{\dots}  \, , \\
    0 & = \Big[ (-1)^{\w_1} \del_{y_1} + (y_3^2 \del_{y_3} + 2 j_3 y_3  )   \Big] \braket{\dots}\, .  
    \end{aligned}
\end{align}
where $\braket{\dots}$ again denotes $\braket{
V^{\w_1}_{j_1}(0,y_1,0)
V^{\w_2}_{j_2}(0,y_2,1)
V^{\w_3}_{j_3}(\infty,y_3,\infty)  }$. In this case, only the first of these equations gets eventually modified away from the collision limit. 

We find that, up to an overall normalization, the odd edge three-point functions read 
\begin{align}
    &\braket{
    V^{\w_1}_{j_1}(0,y_1,0)
    V^{\w_2}_{j_2}(0,y_2,1)
    V^{\w_3}_{j_3}(\infty,y_3,\infty)  }  \\
    &= 
    y_3^{j_1 + j_2 - j_3 - \frac{k}{2}} 
    (1 + (-1)^{\w_1}y_1 y_3 + (-1)^{\w_3}y_2 y_3 )^{\frac{k}{2} - j_1 - j_2 - j_3} \, , \nn 
\end{align}
in the collision limit, thus matching the result in \cite{Dei:2021xgh}, see their Eq.~(5.37c). Moreover, as in the even edge case, we can infer the solution for generic insertion points from the general expression in Eq.~\eqref{oddfinal2}. For this, we set $\w_3 = \w_1+\w_2+1$ and carefully take the limit $a_3 \to 0, \, a_{1,2} \to \infty $ with the products $a_{1}a_3$ and $a_{2}a_3$ fixed. This yields 
\begin{align}
\label{eq: general odd edge}
    &\braket{
    V^{\w_1}_{j_1}(0,y_1,0)
    V^{\w_2}_{j_2}(1,y_2,1)
    V^{\w_3}_{j_3}(\infty,y_3,\infty)  } = y_3^{j_1 + j_2 - j_3 - \frac{k}{2}} \\
    &\times 
    \left(1 + (-1)^{\w_1+1} \frac{(\w_1 + \w_2)!}{\w_1! \w_2!}  y_3 + (-1)^{\w_1}y_1 y_3 + (-1)^{\w_3}y_2 y_3 \right)^{\frac{k}{2} - j_1 - j_2 - j_3} \, . \nn 
\end{align}
One can check that this matches the $y$-dependence given in the conjecture of Eq.~\eqref{3pt-odd-parity} for correlators with appropriate spectral flow assignments. Moreover, upon using Eq.~\eqref{seriesidybasis1} and \eqref{seriesidybasis2} we also see that, as expected, the expressions in Eq.~\eqref{eq: general even solution} and Eq.~\eqref{eq: general odd edge} are related via series identifications.

\subsection{Three-point functions with unflowed insertions}
\label{sec:unflowed}

So far, we have considered three-point functions where all vertex operators had non-zero spectral flow charges. However, it is natural to expect that the above results include the special cases where some of the insertions are unflowed.
We now show how the latter are obtained. Note that we still assume $\w_3 \ge \w_{1,2}$, as in the previous sections.

Let us start by discussing the case of a single unflowed insertion, namely $\w_1=0$. 
The fusion rules in Eq.~\eqref{fusionrules} can then be satisfied iff $\w_3 = \w_2$ or $\w_3 = \w_2 + 1$. These are exactly the two cases that were computed in \cite{Cagnacci:2013ufa} in full generality from $m$-basis techniques. In this sense, the results presented in this section are not new, but we include them for completeness. 
Hence, the relevant correlators correspond to the spectral flow assignments $(0,\w,\w+1)$ and $(0,\w,\w)$. By means of 
\begin{equation}
     V_{j}(x,z) = 
     {\cal{N}}\left(j\right)V_{\frac{k}{2}-j,h=j}^{1} (x,z) = {\cal{N}}(j)
     \lim_{y\to \infty}
     y^{k-2j}
     V_{\frac{k}{2}-j}^{1} (x,y,z) 
     \, ,  
    \label{seriesidentifXbasisW0}
\end{equation}
which is a particular case of \eqref{seriesidybasis1}, these can be obtained from the three-point functions with charges $(1,\w,\w+1)$ and $(1,\w,\w)$, respectively. More precisely, we have 
\begin{equation} 
\langle V_{j_1}
    V_{j_2}^{\w}(y_2)
    V_{j_3}^{\w + 1}(y_3) \rangle =  
    {\cal{N}}(j_1)
     \lim_{y_1\to \infty}
     y_1^{k-2j_1}
     \langle 
     V_{\frac{k}{2}-j_1}^1(y_1)
    V_{j_2}^{\w}(y_2)
    V_{j_3}^{\w + 1}(y_3) \rangle \, ,
    \label{seriesid0ww}
\end{equation}
and 
\begin{equation} 
\langle V_{j_1}
    V_{j_2}^{\w}(y_2)
    V_{j_3}^{\w}(y_3) \rangle =  
    {\cal{N}}(j_1)
     \lim_{y_1\to \infty}
     y_1^{k-2j_1}
     \langle 
     V_{\frac{k}{2}-j_1}^1(y_1)
    V_{j_2}^{\w}(y_2)
    V_{j_3}^{\w}(y_3) \rangle
    \, ,   
    \label{seriesid0ww+1}
\end{equation}
where we have abbreviated $V_{j_1}(0,0)\equiv V_{j_1}$. 

Focusing on \eqref{seriesid0ww}, the RHS involves an even edge correlator. We thus need to consider the appropriate limit of Eq.~\eqref{eq: even edge solution}, which gives
\begin{equation}
\label{eq: oneunflowed1}
    \langle V_{j_1}
    V_{j_2}^{\w}(y_2)
    V_{j_3}^{\w+1}(y_3) \rangle = y_3^{j_1 + j_2 - j_3 - \frac{k}{2}} \left( 1 - y_3 - (-1)^\w y_2 y_3 \right)^{\frac{k}{2} - j_1 - j_2 - j_3} \, . 
\end{equation}
up to an overall constant. On the other hand, the RHS of \eqref{seriesid0ww+1} is obtained as the appropriate limit of the solution in Eq.~\eqref{oddfinal1}, namely
\begin{align}
    & \hspace{-0.2cm}  \langle 
     V_{j_1}^1(y_1)
    V_{j_2}^{\w}(y_2)
    V_{j_3}^{\w}(y_3) \rangle =  
    (y_1 - \w)^{j_2 + j_3 - j_1 - \frac{k}{2}}
    (y_2 + (-1)^\w)^{j_1 + j_3 - j_2 - \frac{k}{2}}
    (y_3 - 1)^{j_1 + j_2 - j_3 - \frac{k}{2}} \nn \\
    & \hspace{-0.2cm} \times 
    \Big( 
    (-1)^{\w+1} (\w+1) + (-1)^{\w} y_1 - y_2 + (-1)^{\w} y_3 - (\w-1) y_2 y_3 + y_1 y_2 y_3
    \Big)^{\frac{k}{2} - j_1 - j_2 - j_3} \, . 
\end{align}
Hence, we get
\begin{equation}
    \langle V_{j_1}
    V_{j_2}^{\w}(y_2)
    V_{j_3}^{\w }(y_3) \rangle = 
    \left( y_2 + (-1)^\w  \right)^{j_3 - j_1 - j_2}
    \left( y_3 - 1 \right)^{j_2 - j_1 - j_3}
    \left(  (-1)^{\w} +  y_2 y_3 \right)^{j_1 - j_2 - j_3} \, . 
\end{equation}
up to the overall constant. Finally, we consider correlators with exactly two unflowed insertions, $\w_1 = \w_2 = 0$. According to \eqref{fusionrules} this can only be non-trivial for $\w_3 = 1$. By using again Eq.~\eqref{seriesidentifXbasisW0} we get  
\begin{align}
    \langle V_{j_1}
    V_{j_2}
    V_{j_3}^{1}(y_3) \rangle & =  
    {\cal{N}}(j_2)
     \lim_{y_2\to \infty}
     y_2^{k-2j_2}
     \langle 
     V_{j_1}
    V_{\frac{k}{2}-j_2}^{1}(y_2)
    V_{j_3}^{1}(y_3) \rangle \nn \\
& = y_3^{j_1 + j_2 - j_3 - \frac{k}{2}} (y_3 - 1)^{\frac{k}{2}-j_1 - j_2 - j_3}
\, , 
\end{align}
where in the last line we have ignored an overall normalization factor. In this way, we match all the corresponding results of \cite{Dei:2021xgh}, where the authors showed that this further reproduces the original computations of \cite{Maldacena:2001km,Cagnacci:2013ufa}.

\subsection{Normalization}
\label{sec: Normalization}

So far we have focused on the dependence of the $y$-basis spectrally flowed correlators on the variables $y_1$, $y_2$ and $y_3$, and shown that it matches precisely the predictions of \cite{Dei:2021xgh}. We now describe how the overall normalizations in Eqs.~\eqref{3pt-odd-parity}-\eqref{consdei} are obtained\footnote{This was already discussed in \cite{Iguri:2022eat}, assuming the $y$-dependence of the correlators was as in \cite{Dei:2021xgh}. We reproduce the argument here for completeness}.  

Once again, the argument relies on the SL(2,$\R$) series identifications. Indeed, identities such as those in Eqs.~\eqref{3mapsCorrs1} and \eqref{3mapsCorrs2} must hold exactly, including the normalization factors. Having fixed the $y$-dependence, we can thus determine the normalizations recursively, starting from the unflowed three-point functions of \cite{Teschner:1999ug,Maldacena:2001km}. For instance, we consider the following identity: 
    \begin{equation}
     \lim_{y_3\rightarrow \infty} y_3^{2j_3} \left\langle V^{\w_1}_{j_1}(y_1) \,  V^{\w_2}_{j_2}(y_2) \, V^{\w_3}_{j_3}(y_3) \right\rangle = 
    {\cal{N}}(j_3)\left\langle V^{\w_1}_{j_1}(y_1) \,  V^{\w_2}_{j_2}(y_2) \, V^{\w_3-1}_{\frac{k}{2}-j_3}(0) \right\rangle\, ,
    \label{identifC}
\end{equation}
which will give us a recursion relation for $C_{\boldsymbol{\w}}(j_1,j_2,j_3)$. Since the latter is independent of the $y_i$, we can set $y_1=y_2=0$. 
Using the $y$-dependence derived above, written as in Eqs.~\eqref{3pt-odd-parity} and \eqref{3pt-even-parity}, one finds that the product of $X_I$ factors on the left- an right-hand sides of \eqref{identifC},  both reduce to either
\begin{equation}
    P_{\boldsymbol{\w}}^{j_1+j_2+j_3-k} \, P_{\boldsymbol{\w}+e_1+e_2}^{j_3-j_1-j_2}
    \,
    P_{\boldsymbol{\w}+e_2-e_3}^{j_1-j_2-j_3}
    \,
    P_{\boldsymbol{\w}+e_1-e_3}^{j_2-j_3-j_1},
\end{equation}
or 
\begin{equation}
    P_{\boldsymbol{\w+e_1+e_2-e_3}}^{\frac{k}{2}-j_1-j_2-j_3} \,     
    P_{\boldsymbol{\w}+e_1}^{-j_1+j_2+j_3-\frac{k}{2}}
    \, 
    P_{\boldsymbol{\w}+e_2}^{j_1-j_2+j_3-\frac{k}{2}}
    \,
    P_{\boldsymbol{\w}-e_3}^{j_1+j_2-j_3-\frac{k}{2}}\, ,
\end{equation}
depending on the overall parity of the spectral flow charges. Consequently, in both cases we find that  Eq.~\eqref{identifC} holds iff
\begin{equation}
    C_{\boldsymbol{\w}}(j_1,j_2,j_3) =
    {\cal{N}}(j_3) C_{\boldsymbol{\w}-e_3}\left(j_1,j_2,\frac{k}{2}-j_3\right). 
\end{equation}
Analogous statements can be derived by shifting the spectral flow charges $\w_1$ and $\w_2$ instead. Moreover, one has the identity 
\begin{equation}
    {\cal{N}}(j_1) C\left(\frac{k}{2}-j_1,j_2,j_3 \right) = 
    {\cal{N}}(j_2) C\left(j_1,\frac{k}{2}-j_2,j_3 \right) = 
    {\cal{N}}(j_3) C\left(j_1,j_2,\frac{k}{2}-j_3 \right) \,.
    \label{B123}
\end{equation}
Since ${\cal{N}}(j) {\cal{N}}(\frac{k}{2}-j) = 1$, it follows that, as stated in \eqref{consdei}, $C_{\boldsymbol{\w}}(j_1,j_2,j_3)$ can only be $C(j_1,j_2,j_3)$, i.e.~the unflowed three-point function, or ${\cal{N}}(j_1) C(\frac{k}{2}-j_1,j_2,j_3)$, depending on the parity of $\w_1 + \w_2 + \w_3$. To be precise, this argument is valid for discrete representations, although we expect that it holds
also for the continuous series by analytic continuation in j \cite{Maldacena:2001km,Zamolodchikov:1995aa}.
This concludes our computation of three-point functions with arbitrary spectral flow charges.


\section{Discussion}
\label{sec: discussion}
In this paper, we have computed the $y$-basis string three-point function in AdS$_3$ involving vertex operators with arbitrary spectral flow charges. This provides a proof for the conjecture put forward recently in \cite{Dei:2021xgh}, thus establishing integral expressions for all (primary) three-point functions of the SL(2,$\R$)-WZW model at level $k$, for all $k >3$. 

The subfamily of (odd parity) correlators for which a holomorphic covering map from the worldsheet to the AdS$_3$ boundary exists had been obtained in \cite{Eberhardt:2019ywk,Dei:2021xgh}. Here we have relied on the general structure of local Ward identities (in their $y$-basis formulation) and made extensive use of the SL(2,$\R$) series identifications, whose importance was recently highlighted in \cite{Iguri:2022eat}. This allowed us to extend the methods based on covering maps to all other non-vanishing correlators, as defined by the fusion rules computed in \cite{Maldacena:2001km}. 

Our strategy can be summarised as follows. We first argued that the differential equations satisfied by all $y$-basis three-point functions must take the form given in Eq.~\eqref{GenEqsEven}. Obtaining the general expressions for these equations for all even parity correlators then reduces to computing all unknown coefficients in \eqref{GenEqsEven}. We have provided the relations among correlators with adjacent spectral flow assignments that follow from SL(2,$\R$) series identifications in Eqs.~\eqref{3mapsCorrs1} and \eqref{3mapsCorrs2}. These provide a considerable number of identities between even and odd parity correlators in the limit where one of the $y$ variables is taken to either zero or infinity. This allowed us to evaluate all relevant coefficients $A_{ij}$, $B_{ij} $ and $C_i$ in closed form, as given in Eqs.~\eqref{AijMatrix}-\eqref{CiMatrix}. The derivation of these 21 coefficients involves solving a total of 60 conditions, of which 39 can be taken as consistency checks. The latter turn out to be satisfied in a highly non-trivial manner, related to the existence of a set of identities relating the behaviour of different covering maps in the vicinity of the insertion points.  

The resulting differential equations satisfied by even parity correlators are provided in Eqs.~\eqref{Eq1even}-\eqref{Eq3even}. These show a striking similarity with the cases of odd total spectral flow, a hallmark of the existence of a more direct derivation by means of \textit{adjacent} covering maps. We leave this for future work. Here we have shown that the general solution to these equations, namely Eq.~\eqref{eq: general even solution}, is compatible with the proposal of \cite{Dei:2021xgh}. 

We have also discussed the so-called edge cases, whose spectral flow assignments saturate the fusion rules in Eqs.~\eqref{conditionsmapwi} and \eqref{wiEvenCases}. Some subtleties arise when trying to apply the general method described above in this context. For these cases we have provided an alternative approach, based on an improved version of the $m$-basis methods \cite{Maldacena:2001km,Cagnacci:2013ufa}. We have then described how to obtain correlators involving unflowed insertions. Finally, we fixed the overall normalization of all $y$-basis three-point functions, following the arguments of \cite{Iguri:2022eat}.

At this point, it is natural to ask if an analogous story holds for four-point functions, which encode crucial dynamical information about the theory. A closed formula for four-point functions in the $y$-basis with arbitrary spectral flow assignments in terms of the corresponding unflowed correlator was conjectured in \cite{Dei:2021yom}.  
Here the situation is more subtle: on top of the four $y_i$ variables, four-point functions also depend non-trivially on the worldsheet and spacetime cross-ratios, and must satisfy the corresponding 
Knizhnik-Zamolodchikov equations. It has been known for some time \cite{Minces:2005nb,Cagnacci:2015pka} that the latter intertwine non-trivially with the recursion relations of the type described in section \ref{sec:recursions}. If the structure put forward in \cite{Dei:2021yom} is correct, its proof is likely to work in two steps. First, one should use arguments similar to those we have considered in this paper to show that solutions to the $y$-basis differential equations associated with the four-point functions consist of various prefactors given by powers of the generalized differences $X_{I}$, defined in Eq.~\eqref{X_I-3pt}, multiplied by an arbitrary function of the so-called generalized cross-ratio. Second, one should prove that this arbitrary function must satisfy the same Knizhnik-Zamolodchikov equation as the corresponding \textit{unflowed} four-point function.  
The extension of the proof for the case of correlation functions involving four spectrally-flowed vertex operators is a work in progress.

\acknowledgments

It is a pleasure to thank Andrea Dei, Gabriele Di Ubaldo, Lorenz Eberhardt, Stefano Massai, Julian H.~Toro, David Turton.
The work of D.B.~is supported by the Royal Society Research Grant RGF\textbackslash R1\textbackslash181019.
The work of N.K.~is supported by the ERC Consolidator Grant 772408-Stringlandscape.

\bibliographystyle{JHEP}
\bibliography{refs}

\end{document}